\documentclass[aps,prd,amsmath,twocolumn,amssymbaps,showpacs]{revtex4-2}
\usepackage{graphicx}  
\usepackage{dcolumn}   
\usepackage{bm}        
\usepackage{amssymb}   
\usepackage{amsmath}   
\usepackage{bbm}
\usepackage{draftcopy}
\usepackage{relsize} 
\usepackage{xfrac}    
\usepackage{slashed}
\usepackage{color}
\usepackage{footnote}
\usepackage{array}
\definecolor{dgreen}{cmyk}{1.,0.,1.,0.2}        
\definecolor{orange}{cmyk}{0.,0.353,1.,0.}    

\usepackage[bookmarks]{hyperref}

\newcommand{\di}{{\rm d}}

\newcommand{\be}{\begin{equation}}
\newcommand{\ee}{\end{equation}}                                                                               
\newcommand{\bea}{\begin{eqnarray}}
\newcommand{\eea}{\end{eqnarray}}

\begin{document}
\title{Moat regimes within a $2+1$ flavor Polyakov-quark-meson model}
\author{Gaoqing Cao}
\address{School of Physics and Astronomy, Sun Yat-sen University, Zhuhai 519088, China}

\date{\today}

\begin{abstract}
To better understand recent predictions on the moat regime of quantum chromodynamics (QCD)  matter, this paper extends the previous work within the two-flavor quark-meson (QM) model to the more realistic $2+1$ flavor Polyakov-quark-meson (PQM) model. Mainly, two effects are further taken into account: strange quark and confinement coded through Polyakov loop. Model parameters are chosen to consistently reproduce the pseudocritical temperature from lattice QCD, $T_{\rm C}\sim 156~ {\rm MeV}$, and the baryon chemical potential at the critical end point (CEP) from FRG-QCD, $\mu_{\rm B(CEP)}\sim 635~ {\rm MeV}$. It is found that the basic features of moat regimes for $\sigma$ and $\pi$ mesons remain similar to those from QM model: Moat regimes cover the region where temperature or baryon chemical potential is large; reentrances occur around the critical baryon chemical potential of chiral transition at zero temperature. Thus, the FRG-QCD results can still not be well understood, especially why the extrapolated CEP should be consistent with the boundaries of moat regimes for $\sigma$ and $\pi$ mesons. Nevertheless, some basic features can be understood qualitatively, and it is consistent that the pole energies are increasing functions of momenta in the whole $T-\mu_{\rm B}$ plane. The moat regime and pole energy of $K$ mesons are also studied with the features similar.
\end{abstract}


\maketitle

\section{Introduction}
The states of QCD matter under extreme conditions is one of the oldest yet full of vigour and vitality topics in high energy nuclear physics. It is also the most fundamental question in order to study any thermodynamic properties of any QCD systems. According to low energy nuclear physics~\cite{Weizsacker:1935}, the first kind of QCD matter can be attributed to nuclear matter which is extended from heavy nuclei and expected to exist in the circumstance with low temperature and large baryon density, such as neutron stars~\cite{Li:2008gp}. Due to the saturation properties, it is well known that the transition from vacuum to symmetric nuclear matter is of first order with the critical baryon chemical potential $\mu_{\rm B}^{\rm c}\sim 923~{\rm MeV}$~\cite{Green:1953zza,Green:1954zzb}. In 1974, T.D. Lee and  G.C. Wick proposed to generate a new QCD state, now known as "quark gluon plasma", by creating a high temperature circumstance through heavy ion collisions~\cite{Lee:1974ma}. From both lattice QCD simulations~\cite{Bellwied:2015rza,HotQCD:2018pds} and experimental explorations~\cite{Andronic:2017pug,STAR:2017sal}, it is clear that the change from quark gluon plasma to hadronic phase is a crossover at low baryon density with the pseudocritical temperature $T_{\rm C}\sim 156~ {\rm MeV}$. 

Since 1974,  the $T-\mu_{\rm B}$ phase diagram has been widely studied in chiral effective models,  but what we are really sure about it are the regions around $\mu_{\rm B}^{\rm c}$ and $T_{\rm C}$. According to model calculations, there could be two important critical end points (CEPs) in between: one for the gas-liquid transition of nuclear matter at lower temperature and the other for the chiral symmetry breaking and restoration at higher temperature. Recently, the existence and location of the latter are under dense investigations from both theoretical and experimental sides in high energy nuclear physics, refer to Refs.~\cite{Luo:2017faz,Luo:2022mtp}. On the theoretical side, functional renormalization approaches~\cite{Fu:2019hdw,Gao:2020fbl,Gunkel:2021oya} predicted the location to be in the region with $\mu_{\rm B}=600\sim650~{\rm MeV}$ and $\mu_{\rm B}/T>4$, roughly consistent with that from holographic QCD~\cite{Cai:2022omk} but lying between those from lattice QCD simulations with complex $\mu_{\rm B}$~\cite{Sorensen:2024mry} and finite-size scaling analysis~\cite{Clarke:2024ugt}. On the experimental side, a lot of facilities worldwide aim at exploring the CEP by decreasing the colliding energy thus approaching the high temperature and high density region, such as RHIC/STAR in USA, FAIR/CBM in Germany, HIAF/CEE in China, and NICA/MPD in Russia~\cite{Luo:2017faz,Luo:2022mtp}. 

Under such a trend, it is also important to explore possible exotic phases around the CEP, since the prediction and detection of CEP could be greatly affected and experiments might reach the higher density side of CEP. According to the literatures, the most relevant ones are quarkyonic matter and moat regime. The quarkyonic matter is a quark-baryon coexisting phase where chiral symmetry is restored but confinement remains~\cite{McLerran:2007qj}, and had been applied to neutron stars~\cite{Fukushima:2015bda,McLerran:2018hbz,Cao:2020byn,Cao:2022inx}. The moat regime of QCD was first discovered by applying functional renormalization group to QCD matter (FRG-QCD)~\cite{Fu:2019hdw}, where the wave function renormalization of pions becomes negative near the CEP thus features spatial modulations for the mesonic correlations. Then, the possible signatures of moat regime are studied in great details for heavy ion collisions~\cite{Pisarski:2021qof,Rennecke:2021ovl,Rennecke:2023xhc}, and it was found that particle numbers and their correlations greatly enhance at nonzero momentum by following the feature of spectral function. 

Recently, FRG-QCD~\cite{Fu:2024rto} and quark-meson model~\cite{Topfel:2024iop} revisited the moat regime by considering both $\sigma$ and $\pi$ correlations and exploring their spectral functions. According to FRG-QCD~\cite{Fu:2024rto}, the moat regimes are induced by particle-hole fluctuations, and the boundaries are consistent with the extrapolated CEP and chiral crossover line therefrom up to around the temperature $T=140~{\rm MeV}$. While, the moat boundaries found in QM model~\cite{Topfel:2024iop} are rather far away from the chiral transition line except the low temperature parts. Specifically, the feature of reentrance was found around $\mu_{\rm B}^{\rm c}$ in the QM model, that is, the system exists the moat regime at a lower temperature and then enters it again at a higher temperature.

This work is devoted to understanding the different features of moat regimes found in FRG-QCD and quark-meson model by adopting the more realistic $2+1$ flavor Polyakov-quark-meson model. The paper is organized as follows. In Sec.~\ref{PQM}, we present the whole formalism: the Lagrangian density is given in Sec.~\ref{Lagrangian}, the gap equations are derived in Sec.~\ref{gap}, and the static and pole energies of mesons are discussed in Sec.~\ref{meson} with the detailed derivations of the polarization functions reserved in Appendices.~\ref{PF} and \ref{SME}. In Sec.~\ref{num}, we show our numerical results: the $T-\mu_{\rm B}$ phase diagram is illustrated together with the CEPs from literatures in Sec.~\ref{phase}, and the features of moat regimes are studied both numerically and analytically in Sec.~\ref{moat}. Finally, we summarize in in Sec.~\ref{summary}.

\section{Polyakov-quark-meson model}\label{PQM}
\subsection{The Lagrangian density}\label{Lagrangian}
In Euclidean space with the metric $g^{\mu\nu}=-\delta_{\mu\nu}$, the Lagrangian density of the renormalizable $2+1$ flavor Polyakov-quark-meson (PQM) model~\cite{Schaefer:2009ui} is given by
\begin{widetext}
\begin{eqnarray}
{\cal L}_{\rm PQM}&=&\bar{\psi}\Big[i\slashed{\partial}\!-\! i\gamma^4\!\!\left(ig_s{\cal A}^4\!+\!Q_{\rm q}\mu_{\rm Q}\!+\!{\mu_{\rm B}\over3}\right)-g_{\rm m}T^a\Big(\sigma_a+i\gamma^5\pi_a\Big)\Big]\psi+{\rm Tr}({\cal D}_\mu^\dagger \phi^{\dagger}{\cal D}^\mu \phi-m^2_\phi\phi^\dagger\phi)\nonumber\\
&&-h_1[{\rm Tr}(\phi^\dagger\phi)]^2-h_2{\rm Tr}[(\phi^\dagger\phi)^2]+\kappa\ [{\rm Det}(\phi^\dagger)+{\rm Det}(\phi)]+c_a\ {\rm Tr}\ T^a(\phi^\dagger+\phi)-V(L,L^*),
\end{eqnarray}
\end{widetext}
where $\psi(x)=(u(x),d(x),s(x))^T$ denotes the three-flavor quark field with the charge matrix $Q_{\rm q}={\rm dia}(q_{\rm u}, q_{\rm d}, q_{\rm s})$, ${\cal A}^4$ denotes the interaction matrix with temporal gluons, and $\phi\equiv T^a(\sigma_a+i\,\pi_a)$ is the scalar-pseudoscalar field matrix. We will establish a more general formalism than that in Ref.~\cite{Schaefer:2009ui} by considering the case with finite temperature $T$, baryon chemical potential $\mu_{\rm B}$, and electric chemical potential $\mu_{\rm Q}$. To conveniently introduce $\mu_{\rm Q}$ into the covariant derivatives for mesons, ${\cal D}_\mu=\partial_\mu-Q_{\rm m} \mu_{\rm Q}\delta_{\mu 4}$ with $Q_{\rm m}$ the electric charge matrix, we choose $\sigma_a+i\,\pi_a$ to be the eigenstates of electric charge with eigenvalues $q_{\rm m}$. Correspondingly, the interaction matrices in flavor space are 
\bea
T^0&=&\sqrt{1\over6}{\bf 1}, \ \ \ \ \ \ \ \ \ \ \ T^a={\lambda^a\over2}\ (a=3,6,7,8), \nonumber\\
T^{1,2}&=&{\lambda^1\mp i\lambda^2\over2\sqrt{2}}, \ \ \ \ \ T^{4,5}={\lambda^4\mp i\lambda^5\over2\sqrt{2}}
\eea
with  $\lambda^a$  the Gell-Mann matrices. In the vacuum, $u$ and $d$ quarks are almost degenerate and there is no pseudoscalar condensate, thus only the components $a=0, 8$ of $c_a$ are nonzero. 

The pure gauge potential $V(L,L^*)$ for Polyakov loop $L\equiv {1\over N_{\rm c}}{\rm Tr}_{\rm c}e^{i\,g_s\int_0^{1/T}dx_4{\cal A}^4}$ would take the form given by Munich group~\cite{Ratti:2005jh}:
\bea
\!\!\!\!{V(L,L^*)\over T^4}\!&=&\!-{a\left({T\over T_0}\right)\over2}|L|^2\!-\!{0.75\over6}\left(L^3\!+\!{L^*}^3\right)\!+\!{7.5\over4}|L|^4,\\
\nonumber\\
&& \!\!\! a(x)=6.75-{1.95\over x}+{2.625\over x^{2}}-{7.44\over x^{3}},\nonumber\\\nonumber
\eea
since it predicts a consistent temperature for chiral transition and deconfinement as from lattice QCD simulations~\cite{Schaefer:2009ui}. Due to the interaction term $\sim {\cal A}^4$ between gluons and quarks, $L$ would affect quark dynamics explicitly. In principle, quarks would also affect gluon dynamics. To account for the feedback, we simply modify $T_0$ according to the corrections of the two-loop $\beta$-function of QCD~\cite{Herbst:2010rf}, that is,
\bea
\!\!\!\! T_0\!=\!T_\tau e^{-{1/\alpha_{\rm s}\over b(N_{\rm f},\mu_{\rm B})}},\
 b(N_{\rm f},\mu_{\rm B})\!=\!{33\!-\!2N_{\rm f}\over6\pi}\!-\!{16N_{\rm f}\over\pi}\!\!\left(\!{\mu_B\over 2T_\tau}\!\right)^2
\eea
with $T_\tau=1.77\ {\rm GeV}, \alpha_{\rm s}=0.304$ and $N_{\rm f}=3$. 

\begin{widetext}
\subsection{The gap equations}\label{gap}
By integrating over the quark degrees of freedom, the Lagrangian can be bosonized as
\begin{eqnarray}
{\cal L}_{\rm PQM}&=&{\rm Tr}\ln\Big[i\slashed{\partial}\!-\! i\gamma^4\!\!\left(ig_s{\cal A}^4\!+\!Q_{\rm q}\mu_{\rm Q}\!+\!{\mu_{\rm B}\over3}\right)-g_{\rm m}T^a\Big(\sigma_a+i\gamma^5\pi_a\Big)\Big]+{\rm Tr}({\cal D}_\mu^\dagger \phi^{\dagger}{\cal D}^\mu \phi-m^2_\phi\phi^\dagger\phi)\nonumber\\
&&-h_1[{\rm Tr}(\phi^\dagger\phi)]^2-h_2{\rm Tr}[(\phi^\dagger\phi)^2]+K\ [{\rm Det}(\phi^\dagger)+{\rm Det}(\phi)]+c_a\ {\rm Tr}\ T^a(\phi^\dagger+\phi)-V(L,L^*).\label{Lc}
\end{eqnarray}
In energy momentum space, the Lagrangian becomes
\begin{eqnarray}
{\cal L}_{\rm PQM}&=&{\rm Tr}\ln\Big[\slashed{k}\!-\! i\gamma^4\!\!\left(ig_s{\cal A}^4\!+\!Q_{\rm q}\mu_{\rm Q}\!+\!{\mu_{\rm B}\over3}\right)-g_{\rm m}T^a\Big(\sigma_a+i\gamma^5\pi_a\Big)\Big]-{\rm Tr}[(i q_4-Q_{\rm m} \mu_{\rm Q})^2+{\bf k}^2+m^2_\phi]\phi^\dagger\phi\nonumber\\
&&-h_1[{\rm Tr}(\phi^\dagger\phi)]^2-h_2{\rm Tr}[(\phi^\dagger\phi)^2]+K\ [{\rm Det}(\phi^\dagger)+{\rm Det}(\phi)]+c_a\ {\rm Tr}\ T^a(\phi^\dagger+\phi)-V(L,L^*),\label{Lk}
\end{eqnarray}
where $\slashed{k}\equiv -\gamma^4k_4-\boldsymbol{\gamma}\cdot{\bf k}$ and the summations over the internal energy momenta should be understood in the interaction terms. In heavy ion collisions, $\mu_{\rm Q}$ is usually smaller than $\mu_{\rm B}$, so we only trivially expect chiral condensates ($\sigma_0, \sigma_8$) and Polyakov loop ($L, L^*$) to be nonzero in the ground state. For simplicity, their expectation values are assumed to be homogeneous over the space; then in mean field approximation, the thermodynamic potential can be derived as~\cite{Schaefer:2009ui}
\bea
\Omega_{\rm PQM}&=&V(L,L^*)+{m^2_\phi}(2\sigma_{\rm l}^2+\sigma_{\rm s}^2)+{h_1}(2\sigma_{\rm l}^2+\sigma_{\rm s}^2)^2+{h_2}(2\sigma_{\rm l}^4+\sigma_{\rm s}^4)-2K \sigma_{\rm l}^2\sigma_{\rm s}-2c_{\rm l}\ \sigma_{\rm l}-c_{\rm s}\ \sigma_{\rm s}\nonumber\\
&&-2T\sum_{\rm f=u,d,s}^{t=\pm}\int_{-\infty}^\infty {\di^3  k\over(2\pi)^3}\ln\left[1\!+\!3L^{(t)}\,H_{\rm f}^{t}\!+\!3L^{(-t)}\,(H_{\rm f}^{t})^2\!+\!(H_{\rm f}^{t})^3\right],\ H_{\rm f}^{t}(E_{\bf k}^{\rm f},\mu_{\rm Q},\mu_{\rm B})\equiv e^{-{1\over T}\left(E_{\bf k}^{\rm f}-t\left(q_{\rm f}\mu_{\rm Q}+{\mu_{\rm B}\over 3}\right)\right)},\label{Omg}
\eea
where $L^{(+)}\equiv L, L^{(-)}\equiv L^*$ and $E_{\rm f}=\sqrt{k^2+m_{\rm f}^2}$ with the dynamical masses $m_{\rm u}=m_{\rm d}=g_{\rm m} \sigma_{\rm l}$ and $m_{\rm s}=g_{\rm m} \sigma_{\rm s}$.  Note that it is more convenient to use the chiral condensates $\sigma_{\rm l}\equiv{1\over 2\sqrt{3}}(\sqrt{2}\sigma_0+\sigma_8)$ and $\sigma_{\rm s}\equiv{1\over2\sqrt{3}}(\sqrt{2}\sigma_0-2\sigma_8)$ instead of $\sigma_0$ and $\sigma_8$ to present the thermodynamic potential, as they are directly related to quark masses. Another very important thing: By following the standard treatment~\cite{Herbst:2010rf}, the divergent vacuum fluctuations of quarks are gotten rid of through renormalizations of the coupling constants in the mesonic sector. Without inducing confusions, we still keep the original denotations for the coupling constants, but keep in mind that they always refer to the renormalized ones in the following.

Usually, the Polyakov loop is taken to be real for simplicity, that is $L=L^*$, then the thermodynamic potential becomes
\bea
\Omega_{\rm PQM}&=&V(L,L)+{m^2_\phi}(2\sigma_{\rm l}^2+\sigma_{\rm s}^2)+{h_1}(2\sigma_{\rm l}^2+\sigma_{\rm s}^2)^2+{h_2}(2\sigma_{\rm l}^4+\sigma_{\rm s}^4)-2K \sigma_{\rm l}^2\sigma_{\rm s}-2c_{\rm l}\ \sigma_{\rm l}-c_{\rm s}\ \sigma_{\rm s}\nonumber\\
&&+2T\sum_{\rm f=u,d,s}^{t=\pm}\int_{-\infty}^\infty {\di^3  k\over(2\pi)^3}\ln F_{\rm f}^{t}(E_{\bf k}^{\rm f},L,\mu_{\rm Q},\mu_{\rm B}),\ F_{\rm f}^{t}(E_{\bf k}^{\rm f},L,\mu_{\rm Q},\mu_{\rm B})\equiv\left[1+3L\, H_{\rm f}^{t}+3L\,(H_{\rm f}^{t})^2+(H_{\rm f}^{t})^3\right]^{-1}.
\eea
And the gap equations follow the extremal conditions $\partial \Omega_{\rm PQM}/\partial X=0,\ (X=L, \sigma_{\rm l}, \sigma_{\rm s})$ as
\bea
\left[-{a\left({T\over T_0}\right)}L-{0.75}L^2+{7.5}L^3\right]T^4-6T\sum_{\rm f=u,d,s}^{t=\pm}\int_{-\infty}^\infty {\di^3  k\over(2\pi)^3}\left[H_{\rm f}^{t}+(H_{\rm f}^{t})^2\right]F_{\rm f}^{t}&=&0,\\
4{m^2_\phi}\sigma_{\rm l}\!+\!8{h_1}(2\sigma_{\rm l}^2\!+\!\sigma_{\rm s}^2)\sigma_{\rm l}\!+\!8{h_2}\sigma_{\rm l}^3\!-\!4K \sigma_{\rm l}\sigma_{\rm s}\!-\!2c_{\rm l}+6\sum_{\rm f=u,d}^{t=\pm}\int_{-\infty}^\infty {\di^3  k\over(2\pi)^3}{g_{\rm m}^2\sigma_{\rm l}\over E_{\bf k}^{\rm f}}\left[L\, H_{\rm f}^{t}+2L\,(H_{\rm f}^{t})^2+(H_{\rm f}^{t})^3\right]F_{\rm f}^{t}&=&0,\label{gapl}\\
2{m^2_\phi}\sigma_{\rm s}\!+\!4{h_1}(2\sigma_{\rm l}^2\!+\!\sigma_{\rm s}^2)\sigma_{\rm s}\!+\!4{h_2}\sigma_{\rm s}^3\!-\!2K \sigma_{\rm l}^2\!-\!c_{\rm s}+6\sum_{t=\pm}\int_{-\infty}^\infty {\di^3  k\over(2\pi)^3}{g_{\rm m}^2\sigma_{\rm s}\over E_{\bf k}^{\rm s}}\left[L\, H_{\rm s}^{t}+2L\,(H_{\rm s}^{t})^2+(H_{\rm s}^{t})^3\right]F_{\rm s}^{t}&=&0.\label{gaps}
\eea
For the case $\mu_{\rm Q}=0$, it is easy to check that Eqs.~\eqref{gapl} and \eqref{gaps} imply $\sigma_{\rm l}=\sigma_{\rm s}$ in the limit $c_{\rm l}=c_{\rm s}$, that is, when $s$ quark is degenerate with the light quarks. Moreover, in the chiral limit $c_{\rm l}=0$, we find a trivial solution $\sigma_{\rm l}=0$ to Eqs.~\eqref{gapl} thus exact chiral symmetry is realized for the light quarks and mesons in the Lagrangian.

\subsection{The static and pole energies of mesons}\label{meson}
Now we are going to explore mesonic spectra by taking Taylor expansions over mesonic fluctuations in Eq.\eqref{Lc} based on the mean field values of $\sigma_{\rm l}, \sigma_{\rm s}$ and $L$. Keep quadratic terms for a few lightest mesons that we are interested in, that is, $\sigma, \pi,$ and $K$ mesons, we have
\begin{eqnarray}
{\cal L}_{\rm PQM}^2&=&-{1\over 2V_4}{\rm Tr}\Big[G(x,x') g_{\rm m}T^a\Big(\hat{\sigma}_a(x')+i\gamma^5\hat{\pi}_a(x')\Big)G(x',x)g_{\rm m}T^{a'}\Big(\hat{\sigma}_{a'}(x)+i\gamma^5\hat{\pi}_{a'}(x)\Big)\Big]-{1\over2}\hat{\sigma}({\partial}_\mu{\partial}^\mu+m^2_\phi)\hat{\sigma}\nonumber\\
&&-{1\over2}\hat{\pi}^0({\partial}_\mu{\partial}^\mu+m^2_\phi)\hat{\pi}^0-\hat{\pi}^-({\cal D}_\mu{\cal D}^\mu+m^2_\phi)\hat{\pi}^+
-\hat{K}^{0*}({\partial}_\mu{\partial}^\mu+m^2_\phi)\hat{K}^0-\hat{K}^-({\cal D}_\mu{\cal D}^\mu+m^2_\phi)\hat{K}^+\nonumber\\
&&-h_1\left[(6\sigma_{\rm l}^2+\sigma_{\rm s}^2)\hat{\sigma}^2+(2\sigma_{\rm l}^2+\sigma_{\rm s}^2)(2\hat{\pi}^+\hat{\pi}^-+(\hat{\pi}^0)^2+2\hat{K}^+\hat{K}^-+2\hat{K}^{0*}\hat{K}^0)\right]-h_2\left[3\sigma_{\rm l}^2\hat{\sigma}^2+\sigma_{\rm l}^2(2\hat{\pi}^+\hat{\pi}^-+(\hat{\pi}^0)^2)\right.\nonumber\\
&&\left.+2(\sigma_{\rm s}^2+\sigma_{\rm l}^2-\sigma_{\rm l}\sigma_{\rm s})(\hat{K}^+\hat{K}^-+\hat{K}^{0*}\hat{K}^0)\right]+\kappa\left[{\sigma_{\rm s}\over2}\hat{\sigma}^2+\sigma_{\rm s}\left(\hat{\pi}^+\hat{\pi}^-+{1\over2}(\hat{\pi}^0)^2\right)+\sigma_{\rm l}(\hat{K}^+\hat{K}^-+\hat{K}^{0*}\hat{K}^0)\right]\label{L2}
\end{eqnarray}
\end{widetext}
with $\hat{\sigma}$ the fluctuation of $\sigma_{\rm l}$ and the full quark propagator 
\bea
&&G(x,x')\equiv{\rm diag}(G_{\rm u}(x,x'), G_{\rm d}(x,x'), G_{\rm s}(x,x'))\nonumber\\
&=&i\ [i\slashed{\partial}\!-\! i\gamma^4\!\!\left(ig_s{\cal A}^4\!+\!Q_{\rm q}\mu_{\rm Q}\!+\!{\mu_{\rm B}\over3}\right)-g_{\rm m}T^a\sigma_a]^{-1}.
\eea
Usually, the expectation value of ${\cal A}^4$ is taken to be a diagonal and traceless constant matrix in color space, that is, $g_s{\cal A}^4= T\ {\rm diag}(p_1, p_2, p_3)$ with $p_1 + p_2 + p_3 = 0$. Then, the quark propagators in flavor and color spaces are, respectively, 
 \bea
\!\!\!\!\! G_{\rm f,c}(x,x')=i[i\slashed{\partial}\!-\! i\gamma^4\!\!\left(ip_cT\!+\!Q_{\rm f}\mu_{\rm Q}\!+\!{\mu_{\rm B}\over3}\right)\!-\!g_{\rm m}\sigma_{\rm f}]^{-1}
 \eea
with ${\rm f=u, d, s; c=1, 2, 3}$, and become
\bea
\!\!\!G_{\rm f,c}(k)=i[-\slashed{k}\!-\! i\gamma^4\!\!\left(ip_cT\!+\!Q_{\rm f}\mu_{\rm Q}\!+\!{\mu_{\rm B}\over3}\right)\!-\!g_{\rm m}\sigma_{\rm f}]^{-1}\label{Gk}
\eea
in energy momentum space. Actually, it is by applying \eqref{Gk} that the expression \eqref{Omg} is derived.

Since the full quark propagator $G(x,x')$ is diagonal in flavor and color spaces, only the ones with $T^{a\dagger}=T^{a'}$ are nonzero for the first term in \eqref{L2}. Then, for $\sigma, \pi$ and $K$ mesons, the corresponding polarization functions should be evaluated according to
\bea
\!\!\Pi_{\hat{\sigma}}\!&=&\!-{g_{\rm m}^2\over 8V_4}\sum_{\rm f=u,d}^{\rm c=r,g,b} {\rm tr}[G_{\rm f,c}(k\!+\!q)G_{\rm f,c}(k)],\\
\!\!\Pi_{\hat{\pi}^0}\!&=&\!-{g_{\rm m}^2\over 8V_4}\sum_{\rm f=u,d}^{\rm c=r,g,b} {\rm tr}[G_{\rm f,c}(k\!+\!q)i\gamma^5G_{\rm f,c}(k)i\gamma^5],\\
\!\! \Pi_{\hat{\pi}^+}\!&=&\!-{g_{\rm m}^2\over 2V_4}\sum^{\rm c=r,g,b} {\rm tr}[G_{\rm u,c}(k\!+\!q)i\gamma^5G_{\rm d,c}(k)i\gamma^5],\\
\!\!\Pi_{\rm \hat{K}^0}\!&=&\!-{g_{\rm m}^2\over 2V_4}\sum^{\rm c=r,g,b} {\rm tr}[G_{\rm s,c}(k\!+\!q)i\gamma^5G_{\rm d,c}(k)i\gamma^5],\\
\!\!\Pi_{\rm \hat{K}^+}\!&=&\!-{g_{\rm m}^2\over 2V_4}\sum^{\rm c=r,g,b} {\rm tr}[G_{\rm u,c}(k\!+\!q)i\gamma^5G_{\rm s,c}(k)i\gamma^5],
\eea
where the traces ${\rm tr}$ are only over Dirac matrices and energy momentum space. The explicit evaluations and renormalizations are given in Appendix.~\ref{PF}. There are usually energy momentum independent and dependent terms in the vacuum parts of the polarization functions: The independent terms are simultaneously gotten rid of by the renormalizations of the coupling constants in the mesonic sector, but the dependent terms is not relevant to that thus should be kept, see \eqref{Pisv} for example. Actually, the latter is essential to unlock the moat boundaries from the transition line of chiral symmetry, see the discussions at the end of Appendix.~\ref{SME}. If the latter is not properly renormalized but regularized by an energy momentum cutoff $\Lambda$, such as done in Nambu--Jona-Lasino model, the quark loops always contribute positively to the wave function renormalizations within the effective range $T, \mu_{\rm B} <\Lambda$, thus no moat regimes can be justified at all.

Eventually, the mesonic propagators, generally denoted by $G_{\rm m}(iq_4-q_{\rm m}\mu_{\rm Q},{\bf q})\ (m=\hat{\sigma}, \hat{\pi}^0, \hat{\pi}^+, \hat{K}^{0}, \hat{K}^+)$, can be obtained as
\bea
G_{\hat{\sigma}}^{-1}(iq_4,{\bf q})&=&q_4^2+{\bf q}^2+\tilde{m}_{\hat{\sigma}}^2+2\Pi_{\hat{\sigma}}(iq_4,{\bf q}),\\
G_{\hat{\pi}^0}^{-1}(iq_4,{\bf q})&=&q_4^2+{\bf q}^2+\tilde{m}_{\hat{\pi}}^2+2\Pi_{\hat{\pi}^0}(iq_4,{\bf q}),\label{pi0}\\
G_{\hat{\pi}^+}^{-1}(iQ_4,{\bf q})&=&Q_4^2+{\bf q}^2+\tilde{m}_{\hat{\pi}}^2+\Pi_{\hat{\pi}^+}(iQ_4,{\bf q}),\label{pip}\\
G_{\hat{K}^0}^{-1}(iq_4,{\bf q})&=&q_4^2+{\bf q}^2+\tilde{m}_{\hat{K}}^2+\Pi_{\rm \hat{K}^0}(iq_4,{\bf q}),\label{K0}\\
G_{\hat{K}^+}^{-1}(iQ_4,{\bf q})&=&Q_4^2+{\bf q}^2+\tilde{m}_{\hat{K}}^2+\Pi_{\rm \hat{K}^+}(iQ_4,{\bf q})\label{Kp}
\eea
with $iQ_4\equiv iq_4-e\mu_{\rm Q}$ and the tree-level effective masses:
\bea
\tilde{m}_{\hat{\sigma}}^2&\equiv& m^2_\phi\!+\!{2h_1}(6\sigma_{\rm l}^2\!+\!\sigma_{\rm s}^2)\!+\!6h_2\sigma_{\rm l}^2\!-\!{\kappa}\sigma_{\rm s},\\
\tilde{m}_{\hat{\pi}}^2&\equiv&m^2_\phi\!+\!{2h_1}(2\sigma_{\rm l}^2\!+\!\sigma_{\rm s}^2)\!+\!2h_2\sigma_{\rm l}^2\!-\!{\kappa}\sigma_{\rm s},\\
\tilde{m}_{\hat{K}}^2&\equiv&m^2_\phi\!+\!{2h_1}(2\sigma_{\rm l}^2\!+\!\sigma_{\rm s}^2)\!+\!2h_2(\sigma_{\rm s}^2\!+\!\sigma_{\rm l}^2\!-\!\sigma_{\rm l}\sigma_{\rm s})\!-\!{\kappa}\sigma_{\rm l}.\nonumber\\
\eea
One note that $G_{\rm m}(iq_4-q_{\rm m}\mu_{\rm Q},{\bf q})$ does not depend on the direction of ${\bf q}$ as the system is isotropic, so we can simply consider $G_{\rm m}(iq_4-q_{\rm m}\mu_{\rm Q},|{\bf q}|)$ instead. For $\mu_{\rm Q}=0$, we can easily see that $\hat{\pi}^0$ is degenerate with $\hat{\pi}^\pm$ by comparing \eqref{pi0} and \eqref{pip}, and $\hat{K}^0$ is degenerate with $\hat{K}^+$ by comparing \eqref{K0} and \eqref{Kp}. However, kaons are not necessarily degenerate with their antiparticles at finite $\mu_{\rm B}$ due to the mass and thus density splitting between $s$ and light quarks. 

To study the moat regimes, the static energies, defined by $E_{\rm m}(|{\bf q}|)\equiv G_{\rm m}^{-1/2}(0,|{\bf q}|)$ in the limit $\mu_{\rm Q}=0$, are relevant, and the boundaries are determined by the sign change of the wave function renormalizations, 
\bea
Z^\bot_{\rm m}\equiv {1\over 2}{\partial^2 E_{\rm m}^2\over\partial |{\bf q}|^2}\Big|_{|{\bf q}|=0},
\eea
see the derivations in Appendix.~\ref{SME}. On the other hand, to study the dynamical effects of mesons, the pole energies are more relevant and could be evaluated self-consistently by requiring $G_{\rm m}^{-1}(-q_0(|{\bf q}|),|{\bf q}|)=0$ for a given $|{\bf q}|$. According to the FRG-QCD calculations~\cite{Fu:2024rto}, the static energies would show minima at finite $|{\bf q}|$ when both $T$ and $\mu_{\rm B}$ are large, but the pole energies are always the smallest at $|{\bf q}|=0$. Such a feature would be checked in a much wider region in this work. Before that, we would like to point out that there is a universal and consistent expression for the static and pole energies when the momentum $|{\bf q}|$ is large, that is, $E_{\rm m}\approx q_0(|{\bf q}|)\approx|{\bf q}|$. It is because the leading contributions of the polarization functions are $\sim\ln|{\bf q}|^2$ for the static case and $|{\bf q}|$-independent for $q_0^2=|{\bf q}|^2+m_{\rm m}^2$, see the discussions at the end of Appendix.~\ref{PF}.

\section{Numerical calculations and discussions}\label{num}
In the following, we will take $\mu_{\rm Q}=0$ and focus on the $T-\mu_{\rm B}$ plane. To perform numerical calculations, the model parameters should be fixed first. As the quark-gluon coupling constant $g_s$ could eventually be absorbed into the definition of the Polyakov loop which is then determined according to the gap equations, there are seven unknown model parameters in the quark-meson sector. The coupling constants $h_2, \kappa, c_0$ and $c_8$ can be fixed as~\cite{Schaefer:2008hk}
\bea
&& h_2=46.5, \kappa=4810\ {\rm MeV}, c_0=(286\ {\rm MeV})^3,  \nonumber\\
&&c_8=-(311\ {\rm MeV})^3
\eea
by utilizing the well determined experimental results for pion decay constant and pseudoscalar meson masses in the vacuum,
\bea
&&f_\pi=92.4\ {\rm MeV}, m_\pi=138\ {\rm MeV}, m_K=496\ {\rm MeV}, \nonumber\\
&&m_{\eta'}=963\ {\rm MeV}, m_\eta=539\ {\rm MeV}.
\eea
The corresponding coupling constants in the linear terms of $\sigma_{\rm l}$ and $\sigma_{\rm s}$ are  $c_{\rm l}=(121\ {\rm MeV})^3$ and $c_{\rm s}=(378\ {\rm MeV})^3$, respectively. 

However, there are quite large biases in the scalar meson masses: According to the Particle Group Data,  the T-matrix pole mass of the lowest energy eigenstate $f(500)$ lies in the range $400\sim 550\ {\rm MeV}$. So we will take the average mass $m_{f(500)}=475\ {\rm MeV}$ as the center value to fix the parameter $g_{\rm m}$ by fitting to the first-principle results, and then the parameters $m_\phi$ and $h_1$ change correspondingly with the value of $m_{f(500)}$ for the given pseudoscalar masses. Note that $f(500)$ is a mixed state of $\hat{\sigma}$ and $\hat{\sigma}_{\rm s}$ due to chiral anomaly~\cite{Schaefer:2008hk}. According to recent lattice QCD simulations, the pseudocritical temperature at zero baryon density is $T_{\rm c}=156\pm 1.5~{\rm MeV}$~\cite{HotQCD:2018pds}; and the FRG-QCD calculations predicted the critical end point to be at $(T_{\rm CEP},\mu_{\rm B(CEP)})=(107, 635)~{\rm MeV}$~\cite{Fu:2019hdw}. Fitting to $T_{\rm c}=156~{\rm MeV}$ and $\mu_{\rm B(CEP)}=635~{\rm MeV}$, the quark-meson coupling constant can be fixed to $g_{\rm m}=5.756$. And the values of $m_\phi$ and $h_1$ are listed in Table.\ref{mh} with respect to $m_{f(500)}$.
\begin{table}[!h]
\centering
\renewcommand\arraystretch{1.5} 
\caption{The model parameters $m_\phi$ and $h_1$.}\label{mh}
\begin{tabular}{|m{2.5cm}<{\centering}||m{2cm}<{\centering}|m{2cm}<{\centering}|}
\hline
 $m_{f(500)} ({\rm MeV})$ & $m_\phi ({\rm MeV})$ & $h_1$ \\
 \hline
400 & 495 & -5.90 \\
\hline
475 & 452& -3.58 \\
\hline
550 & 394&-0.776\\
\hline
\end{tabular}
\end{table}

\subsection{The $T-\mu_{\rm B}$ phase diagram }\label{phase}
\begin{figure}[!htb]
	\begin{center}
		\includegraphics[width=8cm]{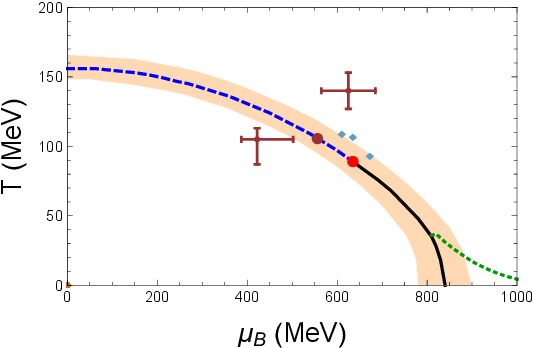}
		\caption{The $T-\mu_{\rm B}$ phase diagram in the Polyakov-quark-meson model with the blue dashed line crossover, the black solid line first-order chiral transition and the red bullet the critical end point at $(635, 89)~{\rm MeV}$. The green dotted line indicates the deconfinement crossover that stars to split from the first-order chiral transition at large $\mu_{\rm B}$. Other theoretical predictions of CEP are also shown for comparison: the cyan diamands are from functional renormalization approaches~\cite{Fu:2019hdw,Gao:2020fbl,Gunkel:2021oya}, and the biased deep-red points are from left to right given by lattice QCD simulations with complex baryon chemical potential~\cite{Sorensen:2024mry}, holographic QCD~\cite{Cai:2022omk}, and finite-size scaling analysis~\cite{Clarke:2024ugt}, respectively.}\label{TmuB}
	\end{center}
\end{figure}

As has mentioned, we take the center value of $f(500)$ mass to be $475\ {\rm MeV}$, then $m_{f(500)}=400, 550\ {\rm MeV}$ can be used to estimate error bars. For the parameters given in  Table.\ref{mh}, the $T-\mu_{\rm B}$ phase diagram can be obtained by solving the gap equations and comparing the thermodynamic potentials between different solutions. The results are illustrated in Fig.~\ref{TmuB} for chiral transition of light quarks and deconfinement together with other theoretical predictions of CEP. 
Note that for crossover, the pseudocritical temperatures are determined by the quickest change points of the order parameters. In the PQM model, the (pseudo)critical temperature of chiral transition is consistent with that of deconfinement up to large $\mu_{\rm B}$, thus agrees with those found for crossover in lattice QCD simulations~\cite{Bellwied:2015rza,HotQCD:2018pds}. As we can see from Fig.~\ref{TmuB}, the transition region is quite consistent the CEPs from functional renormalization approaches~\cite{Fu:2019hdw,Gao:2020fbl,Gunkel:2021oya} and holographic QCD~\cite{Cai:2022omk}, but lies between the CEPs from lattice QCD simulations with complex baryon chemical potential~\cite{Sorensen:2024mry} and finite-size scaling analysis~\cite{Clarke:2024ugt}. 

\begin{figure}[!htb]
	\begin{center}
		\includegraphics[width=8cm]{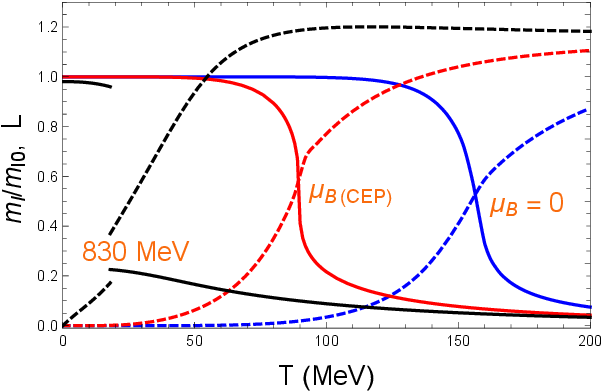}
		\caption{The order parameters, light quark mass $m_{\rm l}$ (reduced by its vacuum value $m_{\rm l0}$, solid lines) and Polyakov loop $L$ (dashed lines) as functions of temperature $T$ for three baryon chemical potentials: $0, \mu_{\rm B (CEP)}$ and $830~{\rm MeV}$.}\label{mlL}
	\end{center}
\end{figure}

Well before the critical baryon chemical potential of chiral transition at zero temperature, $\mu_{\rm B}^{\rm c}\sim 850~{\rm MeV}$, an extra peak begins to show up in the slope of $L$ with respect to $T$, implying splitting between chiral transition and deconfinement. So, there is a region where chiral symmetry is restored but confinement remains at large $\mu_{\rm B}$ and small $T$ --- the so-called "quarkyonic matter" phase. One might notice that the $\mu_{\rm B}^{\rm c}$ found here is even smaller than that of gas-liquid transition of symmetric nuclear matter, it is because the nucleon degrees of freedom are not efficiently taken into account in the PQM model and the supposed gas-liquid transition is missing. Moreover, it is well-known that $L\neq L^*$ at finite $\mu_{\rm B}$, but then the "sign problem" would be involved in chiral effective models~\cite{Fukushima:2006uv}, hence our results for deconfinement can only be understood qualitatively at large $\mu_{\rm B}$. Finally, to demonstrate the occurence of phase transition explicitly, we show the order parameters, light quark mass $m_{\rm l}$ and Polyakov loop $L$, as functions of temperature for different baryon chemical potentials in Fig.~\ref{mlL}
\subsection{The moat regimes}\label{moat}
\begin{figure}[!htb]
	\begin{center}
		\includegraphics[width=8cm]{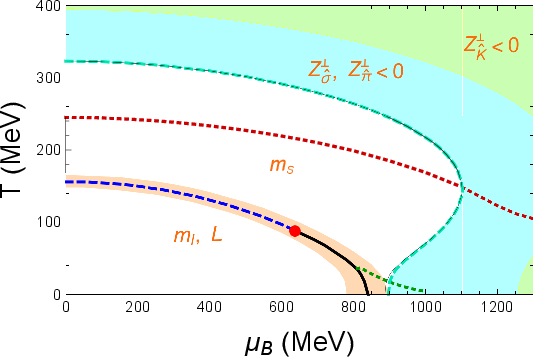}
		\caption{The moat regimes presented by $Z_{\hat{\sigma}}^\bot, Z_{\hat{\pi}}^\bot$ and $Z_{\rm \hat{K}}^\bot<0$ for $\sigma, \pi$ and $K^+$ mesons in the PQM model: cyan for $\sigma$ and $\pi$, and light green for $K^+$. The cyan dashed line and pink dotted line are the corresponding moat boundaries of $\sigma$ and $\pi$, respectively, and the purple dashed and solid lines are those from FRG-QCD calculations. The phase boundaries are also shown with the denotations following Fig.~\ref{TmuB} and the extra deep-red dotted line for the crossover of $s$-quark.}\label{pmoat}
	\end{center}
\end{figure}

The moat regimes for $\sigma, \pi$ and $K^+$ mesons are illustrated together with the phase boundaries in Fig.~\ref{pmoat}. As we can see,  the moat regime for $K^+$ meson occurs at larger $T$ and $\mu_{\rm B}$ than those for $\sigma$ and $\pi$, but the main features are the same. So we will focus on the moat regimes for $\sigma$ and $\pi$ in the following.

Consistent with previous findings~\cite{Fu:2024rto,Topfel:2024iop}, the moat regimes for the chiral partners $\sigma$ and $\pi$ are always in the chiral symmetry restoration phase of light quarks, and due to that they closely overlap with each other. However, at the lower temperature end, it is slightly visible that the moat boundaries of $\pi$ and $\sigma$ split from each other with that of $\pi$ wider. In the following, we try to understand that feature by analyzing the wave function renormalizations given in Eqs. \eqref{Zsigma} and \eqref{Zpi} in the zero temperature limit. 

If we look at their vacuum parts, $Z_{\hat{\sigma}}^\bot$ should be smaller than $Z_{\hat{\pi}}^\bot$ by ${g_{\rm m}^2\over 8\pi^2}\left(1-{m_{\rm l}^2\over m_{\rm l0}^2}\right)$, so the inversion of their actual relative values must come from the thermal part proportional to $m_{\rm l}^2$ in $Z_{\hat{\sigma}}^\bot$. At zero temperature, the system is fully confined with $L=0$, hence $Q_{\rm f}^{-(2)}=0$ in $Z_{\hat{\sigma}}^\bot$ and $Q_{\rm f}^{+(2)}$ can be evaluated explicitly as
\bea
Q_{\rm f}^{+(2)}&\approx& -\int_0^\infty\!\!\!\!{d{k}\over24\pi^2k^2}\left[{1\over E_{{\bf k}}^{\rm f}}\theta\left({\mu_{\rm B}\over 3}-E_{{\bf k}}^{\rm f}\right)-{1\over m_{\rm f}}\right]\nonumber\\
&=&-\int_0^{k_{\rm F}}\!\!\!\!{d{k}\over24\pi^2k^2}\left[{1\over E_{{\bf k}}^{\rm f}}-{1\over m_{\rm f}}\right]+{1\over 24\pi^2m_{\rm f}k_{\rm F}}\nonumber\\
&=&{{\mu_{\rm B}}\over 72\pi^2m_{\rm f}^2k_{\rm F}}
\eea
with $\theta(x)$ the step function and $k_{\rm F}\equiv\sqrt{\left({\mu_{\rm B}\over 3}\right)^2-m_{\rm f}^2}$ the Fermi momentum.  Beyond $\mu_{\rm B}^{\rm c}$, $m_{\rm f}\ll{\mu_{\rm B}\over 3}, m_{\rm l0}$, thus the difference between $Z_{\hat{\sigma}}^\bot$ and $Z_{\hat{\pi}}^\bot$ can be evaluated as
\bea
Z_{\hat{\sigma}}^\bot-Z_{\hat{\pi}}^\bot&=&{g_{\rm m}^2\over 8\pi^2}\left({{\mu_{\rm B}}\over 3k_{\rm F}}-1+{m_{\rm l}^2\over m_{\rm l0}^2}\right)\nonumber\\
&\approx& {g_{\rm m}^2\over 8\pi^2}\left({m_{\rm l}^2\over 2\left({\mu_{\rm B}\over 3}\right)^2}+{m_{\rm l}^2\over m_{\rm l0}^2}\right)>0.
\eea
With respect to that, it is easy to understand why the moat regime of $\pi$ should cover that of $\sigma$ when increasing temperature for a given $\mu_{\rm B}$. In chiral limit, $m_{\rm l}=0$ in the chiral symmetry restoration phase, and the moat boundaries of $\sigma$ and $\pi$ would exactly overlap with each other.

Moreover, with decreasing temperature, we find that the moat boundaries of $\pi$ and $\sigma$ bend back to baryon chemical potentials well beyond $\mu_{\rm B}^{\rm c}$ at zero temperature and are far away from the chiral transition line except the lower temperature end. All the features are qualitatively consistent with those found in QM model~\cite{Topfel:2024iop}, and the former feature actually implies reentrance of moat regimes well beyond $\mu_{\rm B}^{\rm c}$. As has mentioned, the moat boundaries are consistent with the extrapolated CEP and chiral crossover line therefrom up to around $T=140~{\rm MeV}$ in FRG-QCD calculations~\cite{Fu:2024rto}, see the purple lines in Fig.~\ref{pmoat}. Hence, including the effects of strange quark and confinement in the $2+1$ flavor PQM model cannot help to bridge the gap between the predictions of QM model and FRG-QCD. The reason for the failure might be that the mesons are elementary in the QM-like models and do not emerge as quark-antiquark bound states compared to the fundamental QCD theory, thus the correlation between chiral transition and instability of meson fields is weakened and the moat boundaries are not necessarily locked to the chiral transition line.

The reentrance of moat regimes is of course due to the interplay between $T$ and larger $\mu_{\rm B}$, in the following we try to understand that by focusing on $Q_{\rm f}^{t(0)}$ in the wave function renormalization $Z_{\hat{\pi}}^\bot$. 

For most of the reentrance region, the moat boundaries are on top of the deconfinement line, see the green dashed line in Fig.~\ref{pmoat}, so we will take $L=1$ for simplicity and then $Q_{\rm f}^{t(0)}$ is reduced to
\bea
 Q_{\rm f}^{t(0)}&=&-\int_0^\infty{d{k}\over2\pi^2}{1\over E_{{\bf k}}^{\rm f}}{1\over 1+e^{{1\over T}\left(E_{\bf k}^{\rm f}-t{\mu_{\rm B}\over 3}\right)}} \label{Qf0}
 \eea
with $m_{\rm f}\ll{\mu_{\rm B}\over 3}$. As we have pointed out in Appendix.~\ref{SME}, the whole thermal part $\sum_{t=\pm}Q_{\rm f}^{t(0)}$ gives rise to a term $\sim\ln m_{\rm f}$ that exactly cancels the one from vacuum polarization in the limit $m_{\rm f}\rightarrow 0$. To discuss the variation of the thermal part with $T$ for a given $\mu_{\rm B}$, we take the derivative with respective to $T$ and have
 \bea
\!\!\!\!\!\!{\partial\bar{Q}_{\rm f}^{(0)}\over\partial T}\!=\!-{1\over T^2}\int_0^\infty\!\!{d{k}\over2\pi^2}\sum_{t=\pm}{\left(E_{\bf k}^{\rm f}\!-\!t{{\mu}_{\rm B}\over 3}\right)e^{{1\over T}\left(E_{\bf k}^{\rm f}\!-\!t{\mu_{\rm B}\over 3}\right)}\over E_{\bf k}^{\rm f}\left(1\!+\!e^{{1\over T}\left(E_{\bf k}^{\rm f}\!-\!t{\mu_{\rm B}\over 3}\right)}\right)^2}.
\eea
This is convergent in the limit $m_{\rm f}\rightarrow 0$, so we consider such a limit for simplicity and get
 \bea
{\partial\bar{Q}_{\rm f}^{(0)}\over\partial T}\!=\!{H(\tilde{\mu}_{\rm B})\over T}, H(\tilde{\mu}_{\rm B})\!\equiv\!-\!\!\int_0^\infty\!\!\!{d\tilde{k}\over2\pi^2}{1\over \tilde{k}}\sum_{t=\pm}\!\!{\left(\tilde{k}\!-\!t{\tilde{\mu}_{\rm B}\over 3}\right)\!e^{\tilde{k}\!-\!t{\tilde{\mu}_{\rm B}\over 3}}\over \left(1\!+\!e^{\tilde{k}\!-\!t{\tilde{\mu}_{\rm B}\over 3}}\right)^2}\nonumber\\
\eea
with $\tilde{k}\equiv k/T$ and $\tilde{\mu}_{\rm B}\equiv{\mu}_{\rm B}/T$. 

\begin{figure}[!htb]
	\begin{center}
		\includegraphics[width=8cm]{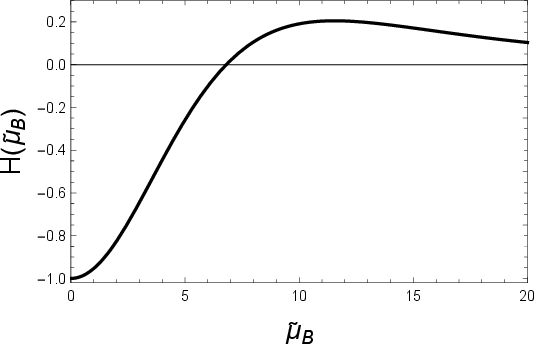}
		\caption{The auxiliary function $H(\tilde{\mu}_{\rm B})$ with $\tilde{\mu}_{\rm B}\equiv{\mu}_{\rm B}/T$.}\label{H}
	\end{center}
\end{figure}

The sign of $H(\tilde{\mu}_{\rm B})$ controls the monotonic behavior of $\bar{Q}_{\rm f}^{(0)}$ with respect to $T$: increasing for $H(\tilde{\mu}_{\rm B})>0$ and decreasing for $H(\tilde{\mu}_{\rm B})<0$. As shown in Fig.~\ref{H}, the sign of $H(\tilde{\mu}_{\rm B})$  changes around $\tilde{\mu}_{\rm B}=7$. 
So start from the moat regime for a large ${\mu}_{\rm B}$ at zero temperature, $\bar{Q}_{\rm f}^{(0)}$  increases with $T$ firstly and then decreases, indicating a peak in $Z_{\hat{\pi}}^\bot$. In the PQM model, the peak of $Z_{\hat{\pi}}^\bot$ is positive well beyond $\mu_{\rm B}^{\rm c}$ thus reentrance is found; for even larger ${\mu}_{\rm B}$, the peak is negative and the whole temperature region is in moat regime. In a word, the reentrance happens because temperature favors moat regime itself on one hand but reduces the chemical potential effect through $\tilde{\mu}_{\rm B}$ on the other hand. For comparison, the baryon number density from light quarks takes the form
\bea
 n_{\rm B}\approx \!\int_0^\infty\!\!{k^2d{k}\over\pi^2}\sum_{t=\pm}\!{t\over 1\!+\!e^{{1\over T}\left(k-t{\mu_{\rm B}\over 3}\right)}}={\mu_{\rm B}\over9}\left[T^2\!+\!\left({\mu_{\rm B}\over3\pi}\right)^2\right]\label{nB}\nonumber\\
\eea 
in the chiral symmetry restoration phase, monotonically increasing with $T$ for a given $\mu_{\rm B}$. Actually, the main difference between \eqref{Qf0} and \eqref{nB} is the order of $k$ in the integrand, and temperature does not suppress chemical potential effect at all in Eq.~\eqref{nB}.

\begin{figure}[!htb]
	\begin{center}
		\includegraphics[width=8cm]{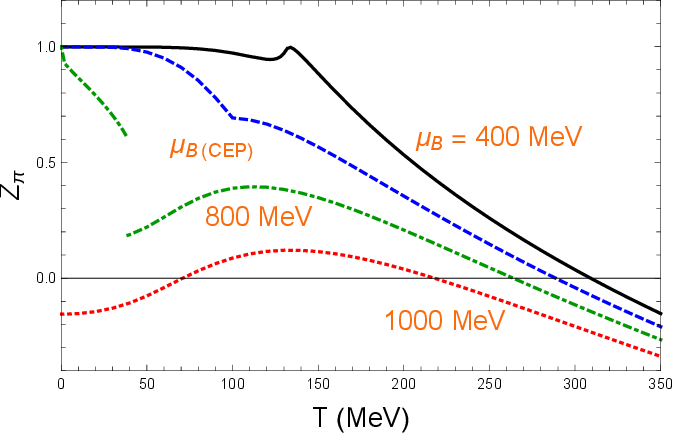}
		\caption{The wave function renormalization of $\pi$ mesons, $Z_{\hat{\pi}}^\bot$, as a function of temperature $T$ for different baryon chemical potentials, ${\mu}_{\rm B}=400, 635~ ( \mu_{\rm B (CEP)}), 800$ and $1000~{\rm MeV}$.}\label{PZpi}
	\end{center}
\end{figure}

We demonstrate the full calculations of $Z_{\hat{\pi}}^\bot$ in Fig.~\ref{PZpi} where the effects of chiral symmetry breaking and restoration and (de-)confinement are taken into account self-consistently.
For the case with ${\mu}_{\rm B}=800~{\rm MeV}$,  we find that $Z_{\hat{\pi}}^\bot$ does not decrease quickly enough across the first-order transition that it is still positive after the transition, that is, the moat boundary is not consistent with the first-order transition. Moreover, there are always local peaks in the curves for all ${\mu}_{\rm B}$ except $ \mu_{\rm B (CEP)}$, but the reasons are quite different. For ${\mu}_{\rm B}< \mu_{\rm B (CEP)}$, the peaks and dips are caused by the chiral crossover with their locations consistent with the pseudocritical temperatures and eventually merging into a point of reflection at $ \mu_{\rm B (CEP)}$.  For ${\mu}_{\rm B}> \mu_{\rm B (CEP)}$, the peaks separate from the chiral transition points and are due to the nonmonotonic feature of $\bar{Q}_{\rm f}^{(0)}$ in the chiral symmetry restoration phase, as has just been discussed. For a small ${\mu}_{\rm B}$, we can only find one zero point for $Z_{\hat{\pi}}^\bot$ because it is shifted to a larger value due to chiral symmetry breaking at the lower temperature end. The latter can be understood more physically: The pseudoGoldstone bosons, pions, dominate the thermodynamics in the chiral symmetry breaking phase. If $Z_{\hat{\pi}}^\bot<0$ there, the sound velocity square would be negative, which means that the pressure unphysically decreases with the energy for a pion gas. For a ${\mu}_{\rm B}$ well beyond $\mu_{\rm B}^{\rm c}$, there are two zero points that can account for the reentrance feature of $Z_{\hat{\pi}}^\bot$. 

In that sense, the moat boundaries of $\pi$ and $\sigma$ mesons found in FRG-QCD calculations~\cite{Fu:2024rto} can be understood in the following way. The lower branches, consistent with chiral crossover line, are of course caused by chiral symmetry restoration where the dips are more significant than ours. And the upper branches correspond to the lower branches of the reentrance region found in our work, where the temperature increases with ${\mu}_{\rm B}$. If the calculations are extended to larger temperature in FRG-QCD, we are supposed to find the top branches of the reentrance region where the temperature decreases with ${\mu}_{\rm B}$. It was pointed out in Ref.~\cite{Fu:2024rto} that the moat regimes are mainly dominated by  particle-particle contributions (or Landau damping), such an observation can be easily justified by look at the pole structure ${1/{\bf q\cdot k}}$ in the integrand of \eqref{Qft}. For particle-antiparticle contributions, the structure ${(E_{{\bf k}+{\bf q/2}}^{\rm f}+E_{{\bf k}- {\bf q/2}}^{\rm f})^{-1}}$ is involved and no pole is found in the limit $|{\bf q}|\rightarrow 0$; while for particle-particle contributions, the structure ${(E_{{\bf k}+{\bf q/2}}^{\rm f}-E_{{\bf k}- {\bf q/2}}^{\rm f})^{-1}}$ is involved and the pole structure is recovered in the limit $|{\bf q}|\rightarrow 0$.

\begin{figure}[!htb]
	\begin{center}
		\includegraphics[width=8cm]{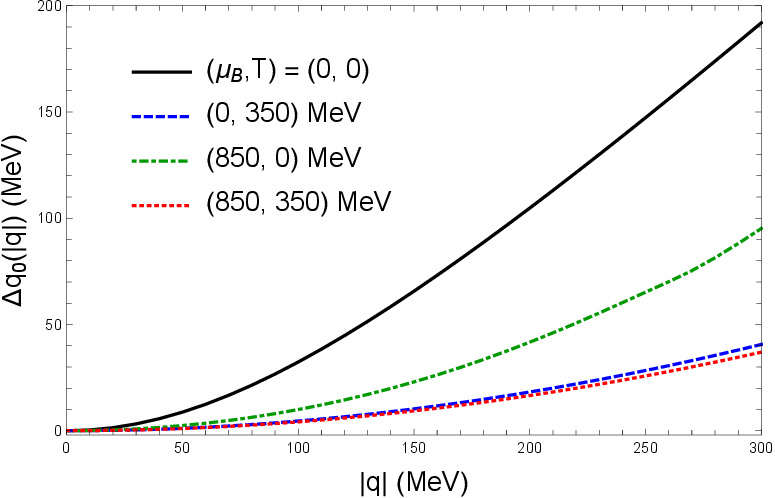}
		\caption{The pole energy $q_0(|{\bf q}|)$ of $\pi$ as a function of momentum $|{\bf q}|$ for different values of temperature and baryon chemical potential.}\label{PE}
	\end{center}
\end{figure}

Finally, we present the pole energy $q_0(|{\bf q}|)$ of $\pi$ as a function of momentum $|{\bf q}|$ in Fig.~\ref{PE} for different values of temperature and baryon chemical potential. 
Consistent with FRG-QCD calculations~\cite{Fu:2024rto}, $q_0(|{\bf q}|)$ is always a monotonically increasing function of $|{\bf q}|$ regardless in the case of vacuum, high density or deep moat regime. Such a feature is also true for $\sigma$ and $K$ mesons, so the pole energy is rather different from the static energy, which could decrease with $|{\bf q}|$ when the wave function renormalization $Z_{m}^\bot$ is negative. Since they are the same at the tree level, the difference must come from the quark loops and the wave function renormalization depends on the energy $q_0$ remarkably. Therefore, to explore dynamical properties in the medium, one should be careful not to take static energy to replace pole energy in the calculations.

\section{Summary}\label{summary}
In this work, the moat regime of QCD matter is explored within the realistic $2+1$ flavor Polyakov-quark-meson model. Compared to the two-flavor quark-meson model, two more effects are further taken into account: strange quark and confinement coded through Polyakov loop. With proper model parameters, the $T-\mu_{\rm B}$ phase diagram obtained is consistent with most predictions on the critical end point. And it is found that the basic features of moat regimes for $\sigma$ and $\pi$ mesons remain similar to those from quark-meson model: They cover the region where temperature or baryon chemical potential is large enough and are basically far away from the chiral transition line; reentrance occurs around the critical baryon chemical potential of chiral transition at zero temperature due to the competition between catalysis and chemical potential suppression effects of temperature. Hence, the effects of strange quark and confinement still cannot help to bridge the gap between the predictions of QM model and FRG-QCD. The reason might be that the mesons are elementary in the QM-like models and do not emerge as quark-antiquark bound states compared to the fundamental QCD theory, thus the correlation between chiral transition and instability of meson fields is weakened and the moat boundaries are not necessarily locked to the chiral transition line, especially the more drastic first-order transition.

Nevertheless, some basic features of the moat regimes found in FRG-QCD calculations can be understood qualitatively:\\ 1. $\sigma$ and $\pi$ mesons are chiral partners, so their moat boundaries are close to each other in the chiral symmetry restoration phase.\\ 2. Since chiral symmetry is not exact in QCD, the terms proportional to quark mass square would enhance the wave function renormalization of $\sigma$ meson at larger temperature or baryon chemical potential, hence its moat boundary lies well within that of pions.\\ 3. The lower branches are caused by chiral symmetry restoration, the upper branches correspond to the lower branches of the reentrance region, and an extra top branches of the reentrance region are expected to exist.

The pole energy is also explored: Consistent with FRG-QCD calculations, it monotonically increases with momentum regardless in the case of vacuum, high density or deep moat regime. Finally, we would like to mention that the moat regime and pole energy of $K$ mesons are also explored and they follow similar features to those of $\sigma$ and $\pi$ mesons. 

\section*{Acknowledgement}
G.C. thanks W.J. Fu from Dalian University of Technology and X.F. Luo from Central China Normal University for helpful discussions. The work is funded by the "Forefront Leading Project of Theoretical Physics" from the National Natural Science Foundation of China with Grant No. 12447102. G.C. is also supported by the Natural Science Foundation of Guangdong Province with Grant No. 2024A1515011225. 

\begin{widetext}
\appendix
\section{Evaluations of polarization functions}\label{PF}
We take $\Pi_{\hat{\sigma}}$ for example, it can be evaluated as the following:
\bea
&&\Pi_{\hat{\sigma}}(iq_4,{\bf q})=\!-{g_{\rm m}^2\over 8}T\sum_{n}\sum_{\rm f=u,d}^{\rm c=r,g,b}\int{d^3{\bf k}\over(2\pi)^3} \!\!{\rm tr}\!\!\left[{i\over -\slashed{k}-\slashed{q}\!-\! i\gamma^4\!\!\left(iq_cT\!+\!Q_{\rm f}\mu_{\rm Q}\!+\!{\mu_{\rm B}\over3}\right)-g_{\rm m}\sigma_{\rm f}}{i\over -\slashed{k}\!-\! i\gamma^4\!\!\left(iq_cT\!+\!Q_{\rm f}\mu_{\rm Q}\!+\!{\mu_{\rm B}\over3}\right)-g_{\rm m}\sigma_{\rm f}}\right]\nonumber\\
&=&\!-{g_{\rm m}^2\over 2}T\sum_{n}\!\!\sum_{\rm f=u,d}^{\rm c=r,g,b}\!\!\int\!{d^3{\bf k}\over(2\pi)^3}\!{ \left(-(\omega_n\!+\!q_4)\!+\!q_cT\!-i Q_{\rm f}\mu_{\rm Q}\!-i {\mu_{\rm B}\over3}\right)\left(-\omega_n\!+\!q_cT\!-i Q_{\rm f}\mu_{\rm Q}\!-i {\mu_{\rm B}\over3}\right)+({\bf k\!+\!q})\cdot{\bf k}-m_{\rm f}^2\over [\left(-(\omega_n\!+\!q_4)\!+\!q_cT\!-i Q_{\rm f}\mu_{\rm Q}\!-i {\mu_{\rm B}\over3}\right)^2\!+\!({\bf k\!+\!q})^2\!+\!m_{\rm f}^2][\left(-\omega_n\!+\!q_cT\!-i Q_{\rm f}\mu_{\rm Q}\!-i {\mu_{\rm B}\over3}\right)^2\!+\!{\bf k}^2\!+\!m_{\rm f}^2]}\nonumber\\
&=&\Pi_{\hat{\sigma}}^{\rm v}-\!{g_{\rm m}^2\over 4}\sum_{\rm f=u,d}^{\rm c=r,g,b}\int{d^3{\bf k}\over(2\pi)^3}\sum_{u,t=\pm}{ -E_{{\bf k}+u {\bf q/2}}^{\rm f}(E_{{\bf k}+u {\bf q/2}}^{\rm f}+u\ t\ i q_4)+{\bf k}^2-{\bf q}^2/4-m_{\rm f}^2\over E_{{\bf k}+u {\bf q/2}}^{\rm f}[(E_{{\bf k}+u {\bf q/2}}^{\rm f}+u\ t\ i q_4)^2-({\bf k}-u\ {\bf q}/2)^2-m_{\rm f}^2]}{1\over 1+e^{[E_{{\bf k}+u {\bf q/2}}^{\rm f}-t(i q_cT+Q_{\rm f}\mu_{\rm Q}+ {\mu_{\rm B}\over3})]/T}}\nonumber\\
&=&\Pi_{\hat{\sigma}}^{\rm v}-\!{3g_{\rm m}^2\over4}\sum_{\rm f=u,d}\int{d^3{\bf k}\over(2\pi)^3}\sum_{u,t=\pm}{ -u\ t\ E_{{\bf k}+u {\bf q/2}}^{\rm f}( i q_4)-u\ {\bf q}\cdot{\bf k}-{\bf q}^2/2-2m_{\rm f}^2\over E_{{\bf k}+u {\bf q/2}}^{\rm f}[2u\ t\ E_{{\bf k}+u {\bf q/2}}^{\rm f}( i q_4)+2u\ {\bf q}\cdot{\bf k}+(i q_4)^2]}\left[LH_{\rm f}^{ut}+2L(H_{\rm f}^{ut})^2+(H_{\rm f}^{ut})^3\right]F_{\rm f}^{ut},
\eea
where $H_{\rm f}^{ut}\equiv H_{\rm f}^{t}(E_{{\bf k}+u {\bf q}/2}^{\rm f},\mu_{\rm Q},\mu_{\rm B})$ and the vacuum term is renormalized in the following way:
\bea
\Pi_{\hat{\sigma}}^{\rm v}(q^2)&\equiv&-{3g_{\rm m}^2\over 2}\sum_{\rm f=u,d}\int{d^4{k}\over(2\pi)^4}{-k\cdot(k+q)-m_{\rm f}^2\over [-k^2+m_{\rm f}^2][-(k+q)^2+m_{\rm f}^2]}\nonumber\\
&=&-{3g_{\rm m}^2\over 2}\sum_{\rm f=u,d}\int{d^4{k}\over(2\pi)^4}{1\over -k^2+m_{\rm f}^2}-{3g_{\rm m}^2\over 4}\sum_{\rm f=u,d}\int{d^4{k}\over(2\pi)^4}{q^2-4m_{\rm f}^2\over [-k^2+m_{\rm f}^2][-(k+q)^2+m_{\rm f}^2]}\nonumber\\
&\Rightarrow&-{3g_{\rm m}^2\over 4}\sum_{\rm f=u,d}\int{d^4{k}\over(2\pi)^4}{q^2-4m_{\rm f}^2\over [-k^2+m_{\rm f}^2][-(k+q)^2+m_{\rm f}^2]}\nonumber\\
&=&{3g_{\rm m}^2\over 2(4\pi)^2}\sum_{\rm f=u,d}(q^2\!-\!4m_{\rm f}^2)\left(\ln{m_{\rm f}\over m_{\rm f0}}+{\sqrt{1\!-\!4{m_{\rm f}^2\over q^2}}}{\rm arccoth}{\sqrt{1\!-\!4{m_{\rm f}^2\over q^2}}}-{\sqrt{1\!-\!4{m_{\rm f0}^2\over q^2}}}{\rm arccoth}{\sqrt{1\!-\!4{m_{\rm f0}^2\over q^2}}}\right)\label{Pisv}
\eea
with $m_{\rm f0}\ ({\rm f=u, d, s})$ the quark masses in vacuum. In the third step, the divergent mass term is simultaneously absorbed by renormalizing the coupling constants in the mesonic sector, as was done when deriving the thermodynamic potential $\Omega_{\rm PQM}$ in \eqref{Omg}. And in the forth step, the divergent integral is renormalized by following dimensional regularization with the condition that it vanishes for all $q^2$ in the vacuum. Similarly, the polarization function of $\hat{\pi}^0$ is
\bea
\Pi_{\hat{\pi}^0}(iq_4,{\bf q})
\!&=&\!\Pi_{\hat{\pi}^0}^{\rm v}\!-\!{3g_{\rm m}^2\over4}\!\!\sum_{\rm f=u,d}\!\!\int\!\!{d^3{\bf k}\over(2\pi)^3}\!\!\sum_{u,t=\pm}\!{ -u\ t\ E_{{\bf k}+u {\bf q/2}}^{\rm f}( i q_4)-u\ {\bf q}\cdot{\bf k}-{\bf q}^2/2\over E_{{\bf k}+u {\bf q/2}}^{\rm f}[2u\ t\ E_{{\bf k}+u {\bf q/2}}^{\rm f}( i q_4)\!+\!2u\ {\bf q}\cdot{\bf k}\!+\!(i q_4)^2]}\!\!\left[LH_{\rm f}^{ut}\!+\!2L(H_{\rm f}^{ut})^2\!+\!(H_{\rm f}^{ut})^3\right]\!F_{\rm f}^{ut},\nonumber\\
\Pi_{\hat{\pi}^0}^{\rm v}(q^2)&\equiv&-{3g_{\rm m}^2\over 2}\sum_{\rm f=u,d}\int{d^4{k}\over(2\pi)^4}{-k\cdot(k+q)+m_{\rm f}^2\over [-k^2+m_{\rm f}^2][-(k+q)^2+m_{\rm f}^2]}\Rightarrow-{3g_{\rm m}^2\over 4}q^2\sum_{\rm f=u,d}\int{d^4{k}\over(2\pi)^4}{1\over [-k^2+m_{\rm f}^2][-(k+q)^2+m_{\rm f}^2]}\nonumber\\
&=&{3g_{\rm m}^2\over 2(4\pi)^2}q^2\sum_{\rm f=u,d}\left(\ln{m_{\rm f}\over m_{\rm f0}}+{\sqrt{1-4{m_{\rm f}^2\over q^2}}}{\rm arccoth}{\sqrt{1-4{m_{\rm f}^2\over q^2}}}-{\sqrt{1\!-\!4{m_{\rm f0}^2\over q^2}}}{\rm arccoth}{\sqrt{1\!-\!4{m_{\rm f0}^2\over q^2}}}\right).
\eea

The polarization functions of $\pi^+, K^0$ and $K^+$ are a little different as the corresponding quark loops involve different flavors. Take $\Pi_{\hat{\pi}^+}$ for example, the polarization function can be evaluated as the following:
\bea
\Pi_{\hat{\pi}^+}(iQ_4,{\bf q})
&=&\Pi_{\hat{\pi}^+}^{\rm v}-{g_{\rm m}^2}\sum^{\rm c=r,g,b}\int{d^3{\bf k}\over(2\pi)^3}\sum_{t=\pm}{ -E_{\bf k}^{\rm d}(E_{\bf k}^{\rm d}-t\  (i Q_4))+({\bf k}+{\bf q})\cdot{\bf k}+m_{\rm u}m_{\rm d}\over E_{\bf k}^{\rm d}[(E_{\bf k}^{\rm d}-t\ (i Q_4))^2-({\bf k}+{\bf q})^2-m_{\rm u}^2]}{1\over 1+e^{[E_{\bf k}^{\rm d}-t(i q_cT+Q_{\rm d}\mu_{\rm Q}+ {\mu_{\rm B}\over3})]/T}}\nonumber\\
&&\ \ \ \ \ \ \ -{g_{\rm m}^2}\sum^{\rm c=r,g,b}\int{d^3{\bf k}\over(2\pi)^3}\sum_{t=\pm}{ -E_{\bf k}^{\rm u}(E_{\bf k}^{\rm u}+t\  (i Q_4))+({\bf k}-{\bf q})\cdot{\bf k}+m_{\rm u}m_{\rm d}\over E_{\bf k}^{\rm u}[(E_{\bf k}^{\rm u}+t\  (i Q_4))^2-({\bf k}-{\bf q})^2-m_{\rm d}^2]}{1\over 1+e^{[E_{\bf k}^{\rm u}-t(i q_cT+Q_{\rm u}\mu_{\rm Q}+ {\mu_{\rm B}\over3})]/T}}\nonumber
\eea
\bea
&=&\Pi_{\hat{\pi}^+}^{\rm v}-{3g_{\rm m}^2}\int{d^3{\bf k}\over(2\pi)^3}\sum_{t=\pm}{ -E_{\bf k}^{\rm d}(E_{\bf k}^{\rm d}-t\  (i Q_4))+({\bf k}+{\bf q})\cdot{\bf k}+m_{\rm u}m_{\rm d}\over E_{\bf k}^{\rm d}[(E_{\bf k}^{\rm d}-t\ (i Q_4))^2-({\bf k}+{\bf q})^2-m_{\rm u}^2]}\left[LH_{\rm d}^{t}+2L(H_{\rm d}^{t})^2+(H_{\rm d}^{t})^3\right]F_{\rm d}^{t}\nonumber\\
&&\ \ \ \ \ \ \ -{3g_{\rm m}^2}\int{d^3{\bf k}\over(2\pi)^3}\sum_{t=\pm}{ -E_{\bf k}^{\rm u}(E_{\bf k}^{\rm u}+t\  (i Q_4))+({\bf k}-{\bf q})\cdot{\bf k}+m_{\rm u}m_{\rm d}\over E_{\bf k}^{\rm u}[(E_{\bf k}^{\rm u}+t\  (i Q_4))^2-({\bf k}-{\bf q})^2-m_{\rm d}^2]}\left[LH_{\rm u}^{t}+2L(H_{\rm u}^{t})^2+(H_{\rm u}^{t})^3\right]F_{\rm u}^{t}\nonumber\\
&=&\Pi_{\hat{\pi}^+}^{\rm v}-{3g_{\rm m}^2}\int{d^3{\bf k}\over(2\pi)^3}\sum_{t=\pm}{ [t\ E_{\bf k}^{\rm d} (i Q_4)+{\bf q}\cdot{\bf k}-m_{\rm d}(m_{\rm d}-m_{\rm u})]\left[LH_{\rm d}^{t}+2L(H_{\rm d}^{t})^2+(H_{\rm d}^{t})^3\right]F_{\rm d}^{t}\over E_{\bf k}^{\rm d}[-2t\ E_{\bf k}^{\rm d} (i Q_4)-2{\bf q}\cdot{\bf k}+ (i Q_4)^2-{\bf q}^2-m_{\rm u}^2+m_{\rm d}^2]}\nonumber\\
&&\ \ \ \ \ \ \ -{3g_{\rm m}^2}\int{d^3{\bf k}\over(2\pi)^3}\sum_{t=\pm}{ [-t\ E_{\bf k}^{\rm u} (i Q_4)-{\bf q}\cdot{\bf k}-m_{\rm u}(m_{\rm u}-m_{\rm d})]\left[LH_{\rm u}^{t}+2L(H_{\rm u}^{t})^2+(H_{\rm u}^{t})^3\right]F_{\rm u}^{t}\over E_{\bf k}^{\rm u}[2t\ E_{\bf k}^{\rm u} (i Q_4)+2{\bf q}\cdot{\bf k}+ (i Q_4)^2-{\bf q}^2-m_{\rm d}^2+m_{\rm u}^2]},
\eea
where the vacuum term is renormalized in the following way:
\bea
\!\!\Pi_{\hat{\pi}^+}^{\rm v}(Q^2)&\equiv&-{6g_{\rm m}^2}\int{d^4{k}\over(2\pi)^4}{-k\cdot(k+Q)+m_{\rm d}m_{\rm u}\over [-k^2+m_{\rm d}^2][-(k+Q)^2+m_{\rm u}^2]}\nonumber\\
&=&-{6g_{\rm m}^2}\int{d^4{k}\over(2\pi)^4}{1\over [-k^2+m_{\rm d}^2]} -{3g_{\rm m}^2}[Q^2-(m_{\rm d}-m_{\rm u})^2]\int{d^4{k}\over(2\pi)^4}{1\over [-k^2+m_{\rm d}^2][-(k+Q)^2+m_{\rm u}^2]}\nonumber\\
&\Rightarrow& -{3g_{\rm m}^2}[Q^2-(m_{\rm d}-m_{\rm u})^2]\int{d^4{k}\over(2\pi)^4}{1\over [-k^2+m_{\rm d}^2][-(k+Q)^2+m_{\rm u}^2]}\nonumber\\
&=&{3g_{\rm m}^2\over (4\pi)^2}[Q^2-(m_{\rm d}-m_{\rm u})^2]\left[\ln{m_{\rm d}m_{\rm u}}+{m_{\rm d}^2-m_{\rm u}^2\over Q^2}\ln{m_{\rm d}\over m_{\rm u}}+M^+_{\rm du}M^-_{\rm du}\ln\left|{M^+_{\rm du}+M^-_{\rm du}\over M^+_{\rm du}-M^-_{\rm du}}\right|-(m_{\rm f}\rightarrow m_{\rm f0})\right]
\eea
with $Q\equiv(Q_4,{\bf q})$ and $M^\pm_{\rm f_1f_2}\equiv \sqrt{1-{(m_{\rm f_1}\pm m_{\rm f_2})^2\over Q^2}}$. In the limit $m_{\rm u}\rightarrow m_{\rm d}$ and $\mu_{\rm Q}=0$, $\Pi_{\hat{\pi}^+}^{\rm v}\rightarrow2\Pi_{\hat{\pi}^0}^{\rm v}$ and the degeneracy between $\hat{\pi}^+$ and $\hat{\pi}^0$ is well reproduced for the vacuum and thermal parts. 

The polarization functions of $K^0$ and $K^+$ can be simply obtained by changing the subscript ${\rm u}$ to ${\rm s}$ and ${\rm d}$ to ${\rm s}$ from $\Pi_{\hat{\pi}^+}$, respectively. We have
\bea
\Pi_{\rm \hat{K}^0}(iq_4,{\bf q})
&=&{3g_{\rm m}^2\over (4\pi)^2}[q^2-(m_{\rm d}-m_{\rm s})^2]\left[\ln{m_{\rm d}m_{\rm s}}+{m_{\rm d}^2-m_{\rm s}^2\over q^2}\ln{m_{\rm d}\over m_{\rm s}}+M^+_{\rm ds}M^-_{\rm ds}\ln\left|{M^+_{\rm ds}+M^-_{\rm ds}\over M^+_{\rm ds}-M^-_{\rm ds}}\right|-(m_{\rm f}\rightarrow m_{\rm f0})\right]
\nonumber\\
&&\!\!\!\!\!\!\!\!\!\!-{3g_{\rm m}^2}\int{d^3{\bf k}\over(2\pi)^3}\sum_{t=\pm}{ t\ E_{\bf k}^{\rm s} (i q_4)+{\bf q}\cdot{\bf k}-m_{\rm s}(m_{\rm s}-m_{\rm d})\over E_{\bf k}^{\rm s}[-2t\ E_{\bf k}^{\rm s} (i q_4)-2{\bf q}\cdot{\bf k}+ (i q_4)^2-{\bf q}^2-m_{\rm d}^2+m_{\rm s}^2]}\left[LH_{\rm s}^{t}+2L(H_{\rm s}^{t})^2+(H_{\rm s}^{t})^3\right]F_{\rm s}^{t}\nonumber\\
&&\!\!\!\!\!\!\!\!\!\!-{3g_{\rm m}^2}\int{d^3{\bf k}\over(2\pi)^3}\sum_{t=\pm}\!{ -t\ E_{\bf k}^{\rm d} (i q_4)-{\bf q}\cdot{\bf k}+m_{\rm d}(m_{\rm s}-m_{\rm d})\over E_{\bf k}^{\rm d}[2t\ E_{\bf k}^{\rm d} (i q_4)+2{\bf q}\cdot{\bf k}+ (i q_4)^2-{\bf q}^2+m_{\rm d}^2-m_{\rm s}^2]}\left[LH_{\rm d}^{t}+2L(H_{\rm d}^{t})^2+(H_{\rm d}^{t})^3\right]F_{\rm d}^{t},\\
\Pi_{\rm \hat{K}^+}(iQ_4,{\bf q})
&=&{3g_{\rm m}^2\over (4\pi)^2}[Q^2-(m_{\rm u}-m_{\rm s})^2]\left[\ln{m_{\rm u}m_{\rm s}}+{m_{\rm u}^2-m_{\rm s}^2\over Q^2}\ln{m_{\rm u}\over m_{\rm s}}+M^+_{\rm us}M^-_{\rm us}\ln\left|{M^+_{\rm us}+M^-_{\rm us}\over M^+_{\rm us}-M^-_{\rm us}}\right|-(m_{\rm f}\rightarrow m_{\rm f0})\right]
\nonumber\\
&&-{3g_{\rm m}^2}\int{d^3{\bf k}\over(2\pi)^3}\sum_{t=\pm}{[ t\ E_{\bf k}^{\rm s} (i Q_4)+{\bf q}\cdot{\bf k}-m_{\rm s}(m_{\rm s}-m_{\rm u})]\left[LH_{\rm s}^{t}+2L(H_{\rm s}^{t})^2+(H_{\rm s}^{t})^3\right]F_{\rm s}^{t}\over E_{\bf k}^{\rm s}[-2t\ E_{\bf k}^{\rm s} (i Q_4)-2{\bf q}\cdot{\bf k}+ (i Q_4)^2-{\bf q}^2-m_{\rm u}^2+m_{\rm s}^2]}\nonumber\\
&&-{3g_{\rm m}^2}\int{d^3{\bf k}\over(2\pi)^3}\sum_{t=\pm}{ [-t\ E_{\bf k}^{\rm u} (i Q_4)-{\bf q}\cdot{\bf k}+m_{\rm u}(m_{\rm s}-m_{\rm u})]\left[LH_{\rm u}^{t}+2L(H_{\rm u}^{t})^2+(H_{\rm u}^{t})^3\right]F_{\rm u}^{t}\over E_{\bf k}^{\rm u}[2t\ E_{\bf k}^{\rm u} (i Q_4)+2{\bf q}\cdot{\bf k}+ (i Q_4)^2-{\bf q}^2+m_{\rm u}^2-m_{\rm s}^2]}.
\eea
Note that as we have set the component quark masses different for $\pi^+, K^0$ and $K^+$ a priori, the momentum integral in one branch can be shifted by ${\bf q}$ without changing the results. However, there are no mass differences in the denominators of the integrands for $\sigma$ and $\pi^0$, thus such a momentum shift could cause ambiguity when one expands the integrand around $|{\bf q}|\sim 0$.

For a large $|{\bf q}|$, the momentum shift is valid for all mesons, so we will take $\hat{\pi}^+$ for example to show the general features of the polarization functions in such a limit. For the static case with $iQ_4=0$, the leading contribution of the vacuum term is $\Pi_{\hat{\pi}^+}^{\rm v}(- |{\bf q}|^2)={3g_{\rm m}^2\over(4\pi)^2}(m_{\rm d0}^2+m_{\rm u0}^2-m_{\rm d}^2-m_{\rm u}^2)\ln |{\bf q}|^2$. And the thermal term vanishes for two facts: The integrals around large $k$ are greatly suppressed due to the presence of $H_{\rm f}^{t}$, and the denominator is of order $o(|{\bf q}|^2)$ compared to the numerator of order $o(|{\bf q}|)$. To study the pole energy, we take the continuation $i Q_4\rightarrow -q_0$, and then we will find the polarization function to be a constant by setting $q_0^2-|{\bf q}|^2=m_{\hat{\pi}^+}^2$, that is,
\bea
\Pi_{\hat{\pi}^+}(m_{\hat{\pi}^+}^2)=\Pi_{\hat{\pi}^+}^{\rm v}(m_{\hat{\pi}^+}^2)+{3g_{\rm m}^2}\int{d^3{\bf k}\over(2\pi)^3}\sum_{\rm f=u,d}^{t=\pm}{1\over 2E_{\bf k}^{\rm f}}\left[LH_{\rm f}^{t}+2L(H_{\rm f}^{t})^2+(H_{\rm f}^{t})^3\right]F_{\rm f}^{t}.
\eea
Note that the thermal term is independent of $m_{\hat{\pi}^+}^2$ and completely controlled by the physical parameters $T, \mu_{\rm B}$ and $\mu_{\rm Q}$. The effective mass $m_{\hat{\pi}^+}$ can be calculated self-consistently from the pole equation, $-m_{\hat{\pi}^+}^2+\tilde{m}_{\hat{\pi}}^2+\Pi_{\hat{\pi}^+}(m_{\hat{\pi}^+}^2)=0$, but is not important in the large $|{\bf q}|$ limit. In summary, the polarization functions of all the mesons show two general features in the large $|{\bf q}|$ limit: proportional to $\ln |{\bf q}|^2$ in the static case and $|{\bf q}|$-independent for $q_0^2-|{\bf q}|^2=m_{\rm m}^2$.

\section{Small momentum expansion of polarization functions}\label{SME}
At finite temperature and density, the system is isotropic, so all the polarization functions should depend on the magnitude rather than the direction of ${\bf q}$. For $\hat{\sigma}$ meson, after setting $iq_4=0$, one can obtain around ${\bf q}^2=0$:
\bea
\Pi_{\hat{\sigma}}&\approx&-{6g_{\rm m}^2\over (4\pi)^2}\sum_{\rm f=u,d}m_{\rm f}^2\ln{m_{\rm f}\over m_{\rm f0}}-{3g_{\rm m}^2\over 2(4\pi)^2}{\bf q}^2\sum_{\rm f=u,d}\left[\ln{m_{\rm f}\over m_{\rm f0}}+{1\over3}\left(1-{m_{\rm f}^2\over m_{\rm f0}}\right)\right]+{3g_{\rm m}^2\over16}\sum_{\rm f=u,d}({\bf q}^2+4m_{\rm f}^2)\sum_{t=\pm}Q_{\rm f}^{t}({\bf q}^2)\nonumber\\
&&+{3g_{\rm m}^2\over4}\sum_{\rm f=u,d}\int{d^3{\bf k}\over(2\pi)^3}\sum_{t=\pm}{ 1\over E_{\bf k}^{\rm f}}\left[LH_{\rm f}^{t}+2L(H_{\rm f}^{t})^2+(H_{\rm f}^{t})^3\right]F_{\rm f}^{t}.
\eea
Here, the momentum expansion of the thermal parts can be evaluated as
\bea
Q_{\rm f}^{t}({\bf q}^2)&\equiv&\int{d^3{\bf k}\over(2\pi)^3}\sum_{u=\pm}{u\over E_{{\bf k}+u {\bf q/2}}^{\rm f}{\bf q}\cdot{\bf k}}\left[LH_{\rm f}^{ut}+2L(H_{\rm f}^{ut})^2+(H_{\rm f}^{ut})^3\right]F_{\rm f}^{ut}\approx Q_{\rm f}^{t(0)}+Q_{\rm f}^{t(2)}{\bf q}^2\label{Qft}
\eea
with the leading and quadratic coefficients
\bea
 Q_{\rm f}^{t(0)}&=&2\int{d^3{\bf k}\over(2\pi)^3}{\partial\over\partial ({\bf k}^2)}{1\over E_{{\bf k}}^{\rm f}}\left[LH_{\rm f}^{t}+2L(H_{\rm f}^{t})^2+(H_{\rm f}^{t})^3\right]F_{\rm f}^{t}=-\int_0^\infty{d{k}\over2\pi^2}{1\over E_{{\bf k}}^{\rm f}}\left[LH_{\rm f}^{t}+2L(H_{\rm f}^{t})^2+(H_{\rm f}^{t})^3\right]F_{\rm f}^{t},\nonumber\\
Q_{\rm f}^{t(2)}&=& \int{d^3{\bf k}\over(2\pi)^3}\left[{1\over2}{\partial^2\over\partial ({\bf k}^2)^2}+(\hat{\bf q}\cdot{\bf k})^2{1\over3}{\partial^3\over\partial ({\bf k}^2)^3}\right]{1\over E_{{\bf k}}^{\rm f}}\left[LH_{\rm f}^{t}+2L(H_{\rm f}^{t})^2+(H_{\rm f}^{t})^3\right]F_{\rm f}^{t}\nonumber\\
&=& \int{d^3{\bf k}\over(2\pi)^3}\left[{1\over2}{\partial^2\over\partial ({\bf k}^2)^2}+{{\bf k}^2\over9}{\partial^3\over\partial ({\bf k}^2)^3}\right]{1\over E_{{\bf k}}^{\rm f}}\left[LH_{\rm f}^{t}+2L(H_{\rm f}^{t})^2+(H_{\rm f}^{t})^3\right]F_{\rm f}^{t}\nonumber\\
&=& \int_0^\infty\!\!{k^2dk\over6\pi^2}{\partial^2\over\partial (k^2)^2}{1\over E_{{\bf k}}^{\rm f}}\!\left[LH_{\rm f}^{t}\!+\!2L(H_{\rm f}^{t})^2\!+\!(H_{\rm f}^{t})^3\right]F_{\rm f}^{t}=\!-\!\int_0^\infty\!\!\!\!{d{k}\over12\pi^2}{\partial\over\partial (k^2)}{1\over E_{{\bf k}}^{\rm f}}\!\left[LH_{\rm f}^{t}\!+\!2L(H_{\rm f}^{t})^2\!+\!(H_{\rm f}^{t})^3\right]F_{\rm f}^{t}\nonumber\\
&=&-\int_0^\infty\!\!\!\!{d{k}\over24\pi^2k^2}\left\{{1\over E_{{\bf k}}^{\rm f}}\!\left[LH_{\rm f}^{t}\!+\!2L(H_{\rm f}^{t})^2\!+\!(H_{\rm f}^{t})^3\right]F_{\rm f}^{t}-({\bf k}\rightarrow0)\right\}.
\eea
That for $\hat{\pi}^0$ follows directly as
\bea
\Pi_{\hat{\pi}^0}&\approx&-{3g_{\rm m}^2\over 2(4\pi)^2}{\bf q}^2\!\!\sum_{\rm f=u,d}\!\ln{m_{\rm f}\over m_{\rm f0}}+{3g_{\rm m}^2\over16}{\bf q}^2\!\!\sum_{\rm f=u,d}\sum_{t=\pm}\!Q_{\rm f}^{t(0)}+{3g_{\rm m}^2\over4}\!\!\sum_{\rm f=u,d}\!\!\int{d^3{\bf k}\over(2\pi)^3}\sum_{t=\pm}\!{ 1\over E_{\bf k}^{\rm f}}\!\!\left[LH_{\rm f}^{t}+2L(H_{\rm f}^{t})^2+(H_{\rm f}^{t})^3\right]F_{\rm f}^{t}.
\eea

Similarly, for $\hat{\pi}^+$, after setting $iQ_4=0$, one can obtain around ${\bf q}^2=0$:
\bea
\Pi_{\hat{\pi}^+}
\!&=&\!-{3g_{\rm m}^2\over (4\pi)^2}\!\left[{\bf q}^2+(m_{\rm d}\!-\!m_{\rm u})^2\right]\!\left[\ln{m_{\rm d}m_{\rm u}}\!+\!{m_{\rm d}^2\!+\!m_{\rm u}^2\over m_{\rm d}^2\!-\!m_{\rm u}^2}\ln{m_{\rm d}\over m_{\rm u}}+{\bf q}^2{m_{\rm d}^4\!-\!m_{\rm u}^4-4m_{\rm d}^2m_{\rm u}^2\ln{m_{\rm d}\over m_{\rm u}}\over2(m_{\rm d}^2\!-\!m_{\rm u}^2)^3}-(m_{\rm f}\rightarrow m_{\rm f0})\right]\nonumber\\
&&+{3g_{\rm m}^2\over2}[{\bf q}^2+(m_{\rm d}-m_{\rm u})^2]\sum_{t=\pm}\sum_{\rm f=u,d}Q_{\rm \hat{\pi}^+f}^{t}({\bf q}^2)+{3g_{\rm m}^2\over2}\sum_{\rm f=u,d}\int{d^3{\bf k}\over(2\pi)^3}\sum_{t=\pm}\!{ 1\over E_{\bf k}^{\rm f}}\left[LH_{\rm f}^{t}+2L(H_{\rm f}^{t})^2+(H_{\rm f}^{t})^3\right]F_{\rm f}^{t}.
\eea
Here, the momentum expansion of the thermal parts can be evaluated as
\bea
Q_{\rm \hat{\pi}^+d}^{t}({\bf q}^2)&=&\int{d^3{\bf k}\over(2\pi)^3}{1\over E_{\bf k}^{\rm d}[-2{\bf q}\cdot{\bf k}-{q}^2-m_{\rm u}^2+m_{\rm d}^2]}\left[LH_{\rm d}^{t}+2L(H_{\rm d}^{t})^2+(H_{\rm d}^{t})^3\right]F_{\rm d}^{t}\nonumber\\
&=&\int_0^\infty{d{k}\over2(2\pi)^2}{1\over E_{\bf k}^{\rm d}}{k\over q}\ln\left|{-2qk+{q}^2+m_{\rm u}^2-m_{\rm d}^2\over 2qk+{q}^2+m_{\rm u}^2-m_{\rm d}^2}\right|\left[LH_{\rm d}^{t}+2L(H_{\rm d}^{t})^2+(H_{\rm d}^{t})^3\right]F_{\rm d}^{t}\approx Q_{\rm \hat{\pi}^+d}^{t(0)}+Q_{\rm \hat{\pi}^+d}^{t(2)}{\bf q}^2;\\
Q_{\rm \hat{\pi}^+u}^{t}({\bf q}^2)&=&\int{d^3{\bf k}\over(2\pi)^3}{1\over E_{\bf k}^{\rm u}[2{\bf q}\cdot{\bf k}-{q}^2+m_{\rm u}^2-m_{\rm d}^2]}\left[LH_{\rm u}^{t}+2L(H_{\rm u}^{t})^2+(H_{\rm u}^{t})^3\right]F_{\rm u}^{t}\nonumber\\
&=&\int_0^\infty{d{k}\over2(2\pi)^2}{1\over E_{\bf k}^{\rm u}}{k\over q}\ln\left|{2qk-{q}^2+m_{\rm u}^2-m_{\rm d}^2\over -2qk-{q}^2+m_{\rm u}^2-m_{\rm d}^2}\right|\left[LH_{\rm u}^{t}+2L(H_{\rm u}^{t})^2+(H_{\rm u}^{t})^3\right]F_{\rm u}^{t}\approx Q_{\rm \hat{\pi}^+u}^{t(0)}+Q_{\rm \hat{\pi}^+u}^{t(2)}{\bf q}^2
\eea
with the leading and quadratic coefficients
\bea
Q_{\rm \hat{\pi}^+d}^{t(0)}&=&-{1\over m_{\rm u}^2-m_{\rm d}^2}\int_0^\infty{2k^2d{k}\over(2\pi)^2}{1\over E_{\bf k}^{\rm d}}\left[LH_{\rm d}^{t}+2L(H_{\rm d}^{t})^2+(H_{\rm d}^{t})^3\right]F_{\rm d}^{t},\\
Q_{\rm \hat{\pi}^+d}^{t(2)}&=&{1\over 3(m_{\rm u}^2-m_{\rm d}^2)^3}\int_0^\infty{2k^2d{k}\over(2\pi)^2}{3(m_{\rm u}^2-m_{\rm d}^2)-4k^2\over E_{\bf k}^{\rm d}}\left[LH_{\rm d}^{t}+2L(H_{\rm d}^{t})^2+(H_{\rm d}^{t})^3\right]F_{\rm d}^{t};\\
Q_{\rm \hat{\pi}^+u}^{t(0)}&=&{1\over m_{\rm u}^2-m_{\rm d}^2}\int_0^\infty{2k^2d{k}\over(2\pi)^2}{1\over E_{\bf k}^{\rm u}}\left[LH_{\rm u}^{t}+2L(H_{\rm u}^{t})^2+(H_{\rm u}^{t})^3\right]F_{\rm u}^{t},\\
Q_{\rm \hat{\pi}^+u}^{t(2)}&=&{1\over 3(m_{\rm u}^2-m_{\rm d}^2)^3}\int_0^\infty{2k^2d{k}\over(2\pi)^2}{3(m_{\rm u}^2-m_{\rm d}^2)+4k^2\over E_{\bf k}^{\rm u}}\left[LH_{\rm u}^{t}+2L(H_{\rm u}^{t})^2+(H_{\rm u}^{t})^3\right]F_{\rm u}^{t}.
\eea
By utilizing partial integral, one can check that 
\bea
\lim_{m_{\rm u}\rightarrow m_{\rm d}}[{\bf q}^2+(m_{\rm d}-m_{\rm u})^2]\sum_{\rm f=u,d}Q_{\rm \hat{\pi}^+f}^{t}({\bf q}^2)={\bf q}^2\lim_{m_{\rm u}\rightarrow m_{\rm d}}\sum_{\rm f=u,d}Q_{\rm \hat{\pi}^+f}^{t(0)}={{\bf q}^2\over 4}\sum_{\rm f=u,d}Q_{\rm f}^{t(0)}
\eea
up to $o({\bf q}^2)$ for $\mu_{\rm Q}=0$, thus $\pi^+$ is degenerate with $\pi^0$ in the limit $m_{\rm u}= m_{\rm d}$ as should be. 

By changing the subscripts, those of $K^0$ and $K^+$ can be directly obtained as
\bea
\!\!\!\!\!\!\!\!\!\!\!\!\!\!\!\!\Pi_{\rm \hat{K}^0}
\!&=&\!-{3g_{\rm m}^2\over (4\pi)^2}\!\left[{\bf q}^2+(m_{\rm d}\!-\!m_{\rm s})^2\right]\!\left[\ln{m_{\rm d}m_{\rm s}}\!+\!{m_{\rm d}^2\!+\!m_{\rm s}^2\over m_{\rm d}^2\!-\!m_{\rm s}^2}\ln{m_{\rm d}\over m_{\rm s}}+{\bf q}^2{m_{\rm d}^4\!-\!m_{\rm s}^4-4m_{\rm d}^2m_{\rm s}^2\ln{m_{\rm d}\over m_{\rm s}}\over2(m_{\rm d}^2\!-\!m_{\rm s}^2)^3}-(m_{\rm f}\rightarrow m_{\rm f0})\right]\nonumber\\
&&+{3g_{\rm m}^2\over2}[{\bf q}^2\!+\!(m_{\rm d}\!-\!m_{\rm s})^2]\sum_{t=\pm}\sum_{\rm f=s,d}[Q_{\rm \hat{K}^0f}^{t(0)}\!+\!Q_{\rm \hat{K}^0f}^{t(2)}{\bf q}^2]\!+\!{3g_{\rm m}^2\over2}\sum_{\rm f=s,d}\int{d^3{\bf k}\over(2\pi)^3}\sum_{t=\pm}{ 1\over E_{\bf k}^{\rm f}}\left[LH_{\rm f}^{t}\!+\!2L(H_{\rm f}^{t})^2\!+\!(H_{\rm f}^{t})^3\right]F_{\rm f}^{t},\\
\!\!\!\!\!\!\!\!\!\!\!\!\!\!\!\!\Pi_{\rm \hat{K}^+}
\!&=&-{3g_{\rm m}^2\over (4\pi)^2}\!\left[{\bf q}^2+(m_{\rm s}\!-\!m_{\rm u})^2\right]\!\left[\ln{m_{\rm s}m_{\rm u}}\!+\!{m_{\rm s}^2\!+\!m_{\rm u}^2\over m_{\rm s}^2\!-\!m_{\rm u}^2}\ln{m_{\rm s}\over m_{\rm u}}+{\bf q}^2{m_{\rm s}^4\!-\!m_{\rm u}^4-4m_{\rm s}^2m_{\rm u}^2\ln{m_{\rm s}\over m_{\rm u}}\over2(m_{\rm s}^2\!-\!m_{\rm u}^2)^3}-(m_{\rm f}\rightarrow m_{\rm f0})\right]\nonumber\\
&&\!+{3g_{\rm m}^2\over2}[{\bf q}^2\!+\!(m_{\rm s}\!-\!m_{\rm u})^2]\sum_{t=\pm}\sum_{\rm f=u,s}[Q_{\rm \hat{K}^+f}^{t(0)}\!+\!Q_{\rm \hat{K}^+f}^{t(2)}{\bf q}^2]\!+\!{3g_{\rm m}^2\over2}\sum_{\rm f=u,s}\int{d^3{\bf k}\over(2\pi)^3}\sum_{t=\pm}{ 1\over E_{\bf k}^{\rm f}}\left[LH_{\rm f}^{t}\!+\!2L(H_{\rm f}^{t})^2\!+\!(H_{\rm f}^{t})^3\right]F_{\rm f}^{t}
\eea
with the leading and quadratic expansion coefficients of the thermal parts
\bea
Q_{\rm \hat{K}^0d}^{t(0)}&=&-{1\over m_{\rm s}^2-m_{\rm d}^2}\int_0^\infty{2k^2d{k}\over(2\pi)^2}{1\over E_{\bf k}^{\rm d}}\left[LH_{\rm d}^{t}+2L(H_{\rm d}^{t})^2+(H_{\rm d}^{t})^3\right]F_{\rm d}^{t},\\
Q_{\rm \hat{K}^0d}^{t(2)}&=&{1\over 3(m_{\rm s}^2-m_{\rm d}^2)^3}\int_0^\infty{2k^2d{k}\over(2\pi)^2}{3(m_{\rm s}^2-m_{\rm d}^2)-4k^2\over E_{\bf k}^{\rm d}}\left[LH_{\rm d}^{t}+2L(H_{\rm d}^{t})^2+(H_{\rm d}^{t})^3\right]F_{\rm d}^{t};\\
Q_{\rm \hat{K}^0s}^{t(0)}&=&{1\over m_{\rm s}^2-m_{\rm d}^2}\int_0^\infty{2k^2d{k}\over(2\pi)^2}{1\over E_{\bf k}^{\rm s}}\left[LH_{\rm s}^{t}+2L(H_{\rm s}^{t})^2+(H_{\rm s}^{t})^3\right]F_{\rm s}^{t},\\
Q_{\rm \hat{K}^0s}^{t(2)}&=&{1\over 3(m_{\rm s}^2-m_{\rm d}^2)^3}\int_0^\infty{2k^2d{k}\over(2\pi)^2}{3(m_{\rm s}^2-m_{\rm d}^2)+4k^2\over E_{\bf k}^{\rm s}}\left[LH_{\rm s}^{t}+2L(H_{\rm s}^{t})^2+(H_{\rm s}^{t})^3\right]F_{\rm s}^{t};\\
Q_{\rm \hat{K}^+s}^{t(0)}&=&-{1\over m_{\rm u}^2-m_{\rm s}^2}\int_0^\infty{2k^2d{k}\over(2\pi)^2}{1\over E_{\bf k}^{\rm s}}\left[LH_{\rm s}^{t}+2L(H_{\rm s}^{t})^2+(H_{\rm s}^{t})^3\right]F_{\rm s}^{t},\\
Q_{\rm \hat{K}^+s}^{t(2)}&=&{1\over 3(m_{\rm u}^2-m_{\rm s}^2)^3}\int_0^\infty{2k^2d{k}\over(2\pi)^2}{3(m_{\rm u}^2-m_{\rm s}^2)-4k^2\over E_{\bf k}^{\rm s}}\left[LH_{\rm s}^{t}+2L(H_{\rm s}^{t})^2+(H_{\rm s}^{t})^3\right]F_{\rm s}^{t};\\
Q_{\rm \hat{K}^+u}^{t(0)}&=&{1\over m_{\rm u}^2-m_{\rm s}^2}\int_0^\infty{2k^2d{k}\over(2\pi)^2}{1\over E_{\bf k}^{\rm u}}\left[LH_{\rm u}^{t}+2L(H_{\rm u}^{t})^2+(H_{\rm u}^{t})^3\right]F_{\rm u}^{t},\\
Q_{\rm \hat{K}^+u}^{t(2)}&=&{1\over 3(m_{\rm u}^2-m_{\rm s}^2)^3}\int_0^\infty{2k^2d{k}\over(2\pi)^2}{3(m_{\rm u}^2-m_{\rm s}^2)+4k^2\over E_{\bf k}^{\rm u}}\left[LH_{\rm u}^{t}+2L(H_{\rm u}^{t})^2+(H_{\rm u}^{t})^3\right]F_{\rm u}^{t}.
\eea

Eventually, for the static case, the wave function renormalizations in front of ${\bf q}^2$ can be summarized as
\bea
Z^\bot_{\hat{\sigma}}&=&1-{3g_{\rm m}^2\over (4\pi)^2}\sum_{\rm f=u,d}\left[\ln{m_{\rm f}\over m_{\rm f0}}+{1\over3}\left(1-{m_{\rm f}^2\over m_{\rm f0}^2}\right)\right]+{3g_{\rm m}^2\over8}\sum_{\rm f=u,d}\sum_{t=\pm}[Q_{\rm f}^{t(0)}+4m_{\rm f}^2Q_{\rm f}^{t(2)}],\label{Zsigma}\\
Z^\bot_{\hat{\pi}^0}&=&1-{3g_{\rm m}^2\over (4\pi)^2}\sum_{\rm f=u,d}\ln{m_{\rm f}\over m_{\rm f0}}+{3g_{\rm m}^2\over8}\sum_{\rm f=u,d}\sum_{t=\pm}Q_{\rm f}^{t(0)},\label{Zpi}\\
Z^\bot_{\hat{\pi}^+}&=&1-{3g_{\rm m}^2\over (4\pi)^2}\left[\ln{m_{\rm d}m_{\rm u}\over m_{\rm d0}m_{\rm u0}}\!+\!{m_{\rm d}^2\!+\!m_{\rm u}^2\over m_{\rm d}^2\!-\!m_{\rm u}^2}\ln{m_{\rm d}\over m_{\rm u}}-{m_{\rm d0}^2\!+\!m_{\rm u0}^2\over m_{\rm d0}^2\!-\!m_{\rm u0}^2}\ln{m_{\rm d0}\over m_{\rm u0}}+(m_{\rm d}-m_{\rm u})^2\left({m_{\rm d}^4\!-\!m_{\rm u}^4-4m_{\rm d}^2m_{\rm u}^2\ln{m_{\rm d}\over m_{\rm u}}\over2(m_{\rm d}^2\!-\!m_{\rm u}^2)^3}\right.\right.\nonumber\\
&&\left.\left. -{m_{\rm d0}^4\!-\!m_{\rm u0}^4-4m_{\rm d0}^2m_{\rm u0}^2\ln{m_{\rm d0}\over m_{\rm u0}}\over2(m_{\rm d0}^2\!-\!m_{\rm u0}^2)^3}\right)\right]+{3g_{\rm m}^2\over2}\sum_{t=\pm}\sum_{\rm f=u,d}[Q_{\rm \hat{\pi}^+f}^{t(0)}+(m_{\rm d}-m_{\rm u})^2Q_{\rm \hat{\pi}^+f}^{t(2)}],\\
Z^\bot_{\rm \hat{K}^0}&=&1-{3g_{\rm m}^2\over (4\pi)^2}\left[\ln{m_{\rm d}m_{\rm s}\over m_{\rm d0}m_{\rm s0}}\!+\!{m_{\rm d}^2\!+\!m_{\rm s}^2\over m_{\rm d}^2\!-\!m_{\rm s}^2}\ln{m_{\rm d}\over m_{\rm s}}-{m_{\rm d0}^2\!+\!m_{\rm s0}^2\over m_{\rm d0}^2\!-\!m_{\rm s0}^2}\ln{m_{\rm d0}\over m_{\rm s0}}+(m_{\rm d}-m_{\rm s})^2\left({m_{\rm d}^4\!-\!m_{\rm s}^4-4m_{\rm d}^2m_{\rm s}^2\ln{m_{\rm d}\over m_{\rm s}}\over2(m_{\rm d}^2\!-\!m_{\rm s}^2)^3}\right.\right.\nonumber\\
&&\left.\left. -{m_{\rm d0}^4\!-\!m_{\rm s0}^4-4m_{\rm d0}^2m_{\rm s0}^2\ln{m_{\rm d0}\over m_{\rm s0}}\over2(m_{\rm d0}^2\!-\!m_{\rm s0}^2)^3}\right)\right]+{3g_{\rm m}^2\over2}\sum_{t=\pm}\sum_{\rm f=s,d}[Q_{\rm \hat{K}^0f}^{t(0)}+(m_{\rm d}-m_{\rm s})^2Q_{\rm \hat{K}^0f}^{t(2)}],\\
Z^\bot_{\rm \hat{K}^+}&=&1-{3g_{\rm m}^2\over (4\pi)^2}\left[\ln{m_{\rm s}m_{\rm u}\over m_{\rm s0}m_{\rm u0}}\!+\!{m_{\rm s}^2\!+\!m_{\rm u}^2\over m_{\rm s}^2\!-\!m_{\rm u}^2}\ln{m_{\rm s}\over m_{\rm u}}-{m_{\rm s0}^2\!+\!m_{\rm u0}^2\over m_{\rm s0}^2\!-\!m_{\rm u0}^2}\ln{m_{\rm s0}\over m_{\rm u0}}+(m_{\rm s}-m_{\rm u})^2\left({m_{\rm s}^4\!-\!m_{\rm u}^4-4m_{\rm s}^2m_{\rm u}^2\ln{m_{\rm s}\over m_{\rm u}}\over2(m_{\rm s}^2\!-\!m_{\rm u}^2)^3}\right.\right.\nonumber\\
&&\left.\left. -{m_{\rm s0}^4\!-\!m_{\rm u0}^4-4m_{\rm s0}^2m_{\rm u0}^2\ln{m_{\rm s0}\over m_{\rm u0}}\over2(m_{\rm s0}^2\!-\!m_{\rm u0}^2)^3}\right)\right].+{3g_{\rm m}^2\over2}\sum_{t=\pm}\sum_{\rm f=u,s}[Q_{\rm \hat{K}^+f}^{t(0)}+(m_{\rm s}-m_{\rm u})^2Q_{\rm \hat{K}^+f}^{t(2)}].
\eea
Note that $Q_{\rm f}^{t(0)}$ would diverge around the infrared integral region $k\sim0$ in the limit $m_{\rm f}\rightarrow0$, and the degree of divergence can be estimated by taking Taylor expansion around $E_{\bf k}^{\rm f}\sim0$ as following
\bea
\sum_{t=\pm}Q_{\rm f}^{t(0)}\sim- \int_0^\Lambda{dk\over2\pi^2}{1\over E_{\bf k}^{\rm f}}\sim {1\over2\pi^2}\log m_{\rm f}.
\eea
Then, in chiral limit, the thermal part would induce a negatively divergent correction to the wave function renormalization when chiral symmetry is fully restored. Thus, the moat boundaries of $\sigma$ and $\pi$ mesons would be exactly the same as the chiral transition line if there are no contributions from vacuum polarizations. By taking the vacuum polarizations into account, one would find that the logarithmic divergences exactly cancel out, and the moat boundaries are not necessarily locked to the chiral transition line.

\end{widetext}

\end{document}